\documentclass[aps,twocolumn,showpacs,showkeys,superscriptaddress]{revtex4-2}

\usepackage[T1]{fontenc}
\usepackage[utf8]{inputenc}
\usepackage{lmodern}
\usepackage{microtype}

\usepackage{mathtools}
\usepackage{amssymb}
\usepackage{bm}
\usepackage{dcolumn}

\def\EE{{\cal E}}
\def\LL{{\cal L}}
\newcommand{\beq}{\begin{equation}}
\newcommand{\eeq}{\end{equation}}
\newcommand{\bea}{\begin{eqnarray}}
\newcommand{\eea}{\end{eqnarray}}

\usepackage{graphicx}
\usepackage[dvipsnames]{xcolor}
\usepackage[normalem]{ulem}
\usepackage{mathrsfs}
\usepackage{orcidlink}
\usepackage{hyperref}
\hypersetup{
  colorlinks=true,
  linkcolor=blue,
  citecolor=cyan,
  urlcolor=MidnightBlue,
  breaklinks=true
}
\usepackage{float}
\begin{document}
\title{Testing charged Simpson–Visser–AdS spacetimes with twin-peak QPOs}
\author{Saeed Ullah Khan \orcidlink{0000-0001-9468-0104}}
\email{Corresponding author: saeedkhan.u@gmail.com}
\affiliation{School of Mathematics and Statistics, Fuzhou University, Fuzhou 350108, Fujian, China}
\author{Javlon Rayimbaev\orcidlink{0000-0001-9293-1838}}
\email{javlonrayimbaev6@gmail.com}
\affiliation{Institute of Theoretical Physics, National University of Uzbekistan, Tashkent 100174, Uzbekistan}
\affiliation{Kimyo International University in Tashkent, Shota Rustaveli Street 156, Tashkent 100121, Uzbekistan}
\affiliation{Tashkent State Technical University, Tashkent 100095, Uzbekistan}
\author{Muhammad Zahid}
\email{zahid.m0011@gmail.com}
\affiliation{Department of Physics, Shanghai University, Shanghai, 200444, China}

\author{Tolaniddin Nurmukhamedov\orcidlink{0000-0002-2507-3674}}\email{ntolaniddin@mail.ru}
\affiliation{Tashkent State Transport University, Tashkent 100067, Uzbekistan}

\author{Abat Muratov\orcidlink{0000-0001-6440-9460}}
\email{acmuratov@mail.ru}
\affiliation{Karakalpak State University, Sh. Abdirov 1, Nukus 230112, Uzbekistan}

\author{Weiwei Wang\orcidlink{0000-0002-9316-480X}}\email{Corresponding author: pde_fzu@163.com}
\affiliation{School of Mathematics and Statistics, Fuzhou University, Fuzhou 350108, Fujian, China}
\date{\today}

\begin{abstract}
In this manuscript, we study circular geodesics and twin-peak high-frequency quasi-periodic oscillations (HF-QPOs) in the charged Simpson--Visser--anti-de Sitter (AdS) spacetime, which depends on a charge parameter $Q$, a regularization scale $b$ and an AdS radius $l$. Written in terms of the areal radius $R=\sqrt{r^{2}+b^{2}}$, the specific energy, angular momentum, Keplerian frequency and innermost stable circular orbit (ISCO) of test particles coincide with those of the Reissner--Nordstr\"om--AdS black hole, so $b$ enters the epicyclic spectrum only through a factor $1-b^{2}/R^{2}$ in the radial frequency. We give this frequency and the ISCO condition in closed form. We then fit the relativistic-precession model to the HF-QPO pairs of GRO~J1655$-$40 and XTE~J1550$-$564, with their dynamical masses included as Gaussian priors. The Schwarzschild limit gives $\chi^{2}=32.9$ for two degrees of freedom. One additional deformation removes this tension: $Q/M=0.814^{+0.059}_{-0.074}$ at fixed $l=100M$, or $l=40.6^{+6.2}_{-4.8}\,M$ at $Q=0$, both with $\chi^{2}\simeq0.13$. These two solutions lie on a single degeneracy valley in the $(Q,l)$ plane, and, at $l=100M$, $b$ is left unconstrained up to $b\simeq3.9M$ (68\%), a limit set by the charge prior rather than by the data. A Kerr black hole with spin $a\simeq0.3$--$0.4$ reproduces the same pairs, so the inferred deformation most likely absorbs the spin neglected by the static metric. Our results therefore describe degeneracies among orbital-frequency deformations; they are not a measurement of an electric charge or of a cosmological constant.
\end{abstract}
\maketitle
\sloppy
\section{Introduction}\label{Sec:introduction}
Theories of gravity provide the fundamental framework for understanding the origin, evolution, and large-scale dynamics of the universe, as well as the behavior of compact relativistic objects, including neutron stars and black holes.
In 1915, Einstein introduced the general theory of relativity (GR), extending special relativity to non-inertial frames and describing gravity as a manifestation of spacetime curvature. GR has passed weak-field tests in the solar system and with gravitational lensing~\cite{Will2014LRR....17....4W,Amendola2008JCAP...04..013A}, and strong-field tests through the detection of gravitational waves~\cite{Abbott2016PhRvL.116f1102A,Pitkin2011LRR.14..5P,Liu2024PhRvD.109h4074L} and the horizon-scale images of the supermassive black holes Sgr A*~\cite{EHT2022ApJ...930L..12E,EHT2022ApJ...930L..17E} and M87*~\cite{EHT2019ApJ...875L...1E,2023MNRAS.523..375S,EHT2023ApJ...957L..20E}.

Despite these successes, GR has well-known theoretical limitations, most notably the spacetime singularities at which curvature and energy density diverge, and the lack of a consistent quantum completion. These open problems motivate modified gravity models and alternative gravitational theories. However, parameter degeneracies often make it difficult to distinguish between different theoretical formulations from astrophysical signatures alone~\cite{2021JCAP...04..082M,2023MNRAS.523..375S}. To address this, model-independent parameterizations of spherically and axially symmetric spacetimes have been developed~\cite{Rezzolla2014PhRvD,Konoplya2020PhRvD.101l4004K}.

In Ref.~\cite{Simpson2019JCAP...02..042S}, Simpson and Visser proposed a simple way of removing the central singularity. Their spherically symmetric metric, and its rotating generalization~\cite{2021JCAP...04..082M}, describes a regular black hole or a traversable wormhole, depending on the value of a regularization length $b$. This parameter may represent quantum-gravity effects or other, as yet unidentified, physics. In the wormhole case the geometry connects two asymptotic regions through a throat, much like a Morris--Thorne wormhole. A traversable wormhole of this kind usually requires exotic matter that violates the null and weak energy conditions in the throat region. Indeed, Simpson--Visser geometries can be obtained as exact solutions of GR sourced by a phantom scalar field together with nonlinear electrodynamics~\cite{Bronnikov2022PhRvD.105d4039B}, which links them naturally to the phantom fields studied in the dark-energy context. Wormholes without exotic matter have also been proposed, such as the Einstein--Dirac--Maxwell wormholes~\cite{2021PhRvL.126j1102B,2022PhRvL.128i1104K,2022EPJC...82..533B}. The regularization has been applied to charged black holes in Ref.~\cite{2021JCAP...07..036F} and, more recently, to charged anti-de Sitter (AdS) black holes, whose geodesics and thermodynamics were studied in Ref.~\cite{Ahmed2026EPJC...86..658A}. Khan et al.~\cite{Khan2026ChJPh.102..711K} also constructed a rotating counterpart of the Simpson--Visser black hole in a modified theory of gravity.

Any modification of GR must be tested against observation, and test-particle motion is one of the most direct ways to do so. The dynamics of massive and massless particles in a given metric carries much of the information about its physical properties~\cite{Chandrasekhar1983mtbh.book.C,2024CQGra..41a5032S,Bambi2017bhlt.book..B}, and X-ray observations of accreting sources probe exactly this region~\cite{Wilkins2012MNRAS}. The motion of neutral and charged test particles near black holes has accordingly been studied in many settings~\cite{Aliev2002MNRAS.336..241A,Vrba2020PhRvD.101l4039V,Zahid2022EPJC...82..494Z,2022ChJPh..78..141K,khan2024circular,2023EPJC...83..854V}.

The Rossi X-ray Timing Explorer (RXTE) produced a large set of timing observations of black-hole transients~\cite{1993ARA&A..31...93V,2012MNRAS.426.1701B,Remillard2006ARA&A..44...49R}. Quasi-periodic oscillations (QPOs) in the X-ray light curves of stellar-mass black-hole binaries are still studied intensively, because their frequencies are set close to the black hole and scale inversely with its mass~\cite{Ingram2019NewAR..8501524I,2022CQGra..39g5021R}. The observed QPOs range from a few mHz to $\sim0.5$~kHz and fall into two broad groups: low-frequency (LF) QPOs, below $\sim30$~Hz, and high-frequency (HF) QPOs, from $\sim40$~Hz up to $\sim450$~Hz~\cite{2012MNRAS.426.1701B}.

HF-QPOs have been used to estimate the masses and spins of black holes~\cite{2001A&A...374L..19A,2012JCAP...09..014B,Motta2014MNRAS.437.2554M,Motta2014MNRAS.439L..65M,2026JCAP...09..062R} and the masses and radii of neutron stars~\cite{1998ApJ...499L..37M,1998ApJ...509L..37K}. Since their frequencies are comparable to the orbital and epicyclic frequencies of matter near the ISCO, they have received considerable theoretical attention~\cite{Jumaniyozov2024EPJC964y,2026EPJC...86..510S}. Most recently, Khan et al.~\cite{Khan2026EPJC...86..597K} studied QPOs around a Kerr-like rotating Simpson--Visser black hole in MOG, and QPOs in Simpson--Visser and other modified-gravity black holes have been examined in Refs.~\cite{2025EPJC...85.1029N,Stuchlik2021JCAP...11..059S,2025PhRvD.112l4018D,Li2026arXiv260214458L}. In the uncharged, asymptotically flat Simpson--Visser metric the twin-peak frequencies $(\nu_{U},\nu_{L})$ are controlled by $b$ alone. Adding $Q$ and $l$ introduces two further deformations acting on the same pair of frequencies, and it becomes a real question which combinations of $(Q,b,l)$ the data can distinguish.

In the present work, we briefly discuss the charged Simpson--Visser--AdS geometry, which were previously analyzed for thermodynamics properties by Ahmed, Al-Badawi, and Fathi~\cite{Ahmed2026EPJC...86..658A}. Our contribution is to further investigate its geometry in terms of its geometrical features, the calculation of the orbital and epicyclic frequencies, the comparison of several twin-peak QPO prescriptions, and a statistical study of which combinations of the deformation parameters can be identified from current HF-QPO measurements. Because $Q$, $b$ and $l$ produce partly degenerate changes in the frequency spectrum, the observational analysis is designed to test identifiability rather than to claim a unique measurement of any single parameter.

The manuscript is organized as follows. In Sec.~\ref{Sec:Charged-SV-AdS}, we briefly discuss the charged Simpson--Visser--AdS spacetime and its horizon structure. In Sec.~\ref{sec3: circular motion} we study circular motion: the effective potential, the specific energy and angular momentum, and the ISCO. Sec.~\ref{sec4:Fund_frequencies and QPOs} is devoted to the fundamental frequencies, the twin-peak QPO models, and the fits to GRO~J1655$-$40 and XTE~J1550$-$564. We summarize our results in Sec.~\ref{sec:conclusion}.
\section{Charged Simpson-Visser-AdS spacetime}\label{Sec:Charged-SV-AdS}
Following Simpson and Visser regularization approach~\cite{Simpson2019JCAP...02..042S}, a singular black-hole metric can be made regular everywhere by the replacement $r^2\to r^2+b^2$~\cite{2021JCAP...07..036F}. This extends the domain of $r$ from $r \in (0, +\infty)$ to $r \in (-\infty, +\infty)$ whenever $b\neq0$, with a minimal two-sphere (a throat or a bounce) at $r=0$. The parameter $b$ is called the regularization parameter. Applied to the Schwarzschild metric, the procedure gives the static Simpson--Visser spacetime, whose line element in the coordinates $x^\mu = (t, r, \theta, \phi)$ reads
\begin{equation}\label{metric}
ds^2=-f(r)dt^2+\frac{1}{f(r)}dr^2+(r^2+b^2)d\Omega^2\ ,
\end{equation}
with  $d\Omega^2 = d\theta^2 + \sin^2\theta\, d\phi^2$, and
\begin{equation}\label{metric1}
f(r)=1 - \frac{2M}{\sqrt{r^2 + b^2}}.
\end{equation}
Here $M$ is the ADM mass. The event horizon of the above Simpson-Visser black holes may be determined by solving $f(r)=0$, as follows: 
\begin{equation}
{r_h}= \pm \sqrt{(2M)^2-b^2}.
\end{equation}
Thus $b<2M$ gives a regular black hole (a ``black bounce''), $b=2M$ a one-way wormhole with an extremal null throat, and $b>2M$ a two-way traversable wormhole~\cite{Simpson2019JCAP...02..042S}.

The Reissner--Nordstr\"om--AdS (RN-AdS) black hole generalizes the Schwarzschild--AdS solution to a non-zero electric charge $Q$, with four-potential $A_{\mu} = - Q \delta_{\mu}^t/{r}$ and lapse function
\begin{eqnarray}\label{function-RN-AdS}
 f(r)=1-\frac{2M}{r}+\frac{Q^2}{r^2}+\frac{r^2}{l^2}.
\end{eqnarray}
Here, $l$ denotes the AdS radius, related to the cosmological constant by $\Lambda =- {3}/{l^2}$.
Applying the same replacement $r^2\to r^2+b^2$~\cite{2021JCAP...07..036F,Ahmed2026EPJC...86..658A} gives the charged Simpson--Visser--AdS spacetime, i.e., the metric~\eqref{metric} with
\begin{equation}\label{function_f}
f(r)=1 - \frac{2M}{\sqrt{r^2 + b^2}} +\frac{Q^2}{r^2+b^2} + \frac{r^2 + b^2}{l^2},
\end{equation}
and four-potential $A_{\mu} = - Q \delta_{\mu}^t/\sqrt{r^2+b^2}$. We consider only neutral test particles below, so $A_\mu$ does not enter the dynamics. We also do not construct the matter source of Eq.~\eqref{function_f}; following common practice, we treat it as a phenomenological geometry. For $l\to\infty$, $M$ is the ADM mass; for finite $l$ we simply call it the mass parameter.

The metric~\eqref{metric} with~\eqref{function_f} reduces to the Schwarzschild--AdS spacetime for $b=Q=0$ and to the RN-AdS spacetime for $b=0$. These limiting cases verify that our charged Simpson–Visser-AdS regular metric interpolates smoothly between the well-known solutions.

It is useful to introduce the areal radius
\begin{equation}\label{eq:Rdef}
R\equiv\sqrt{r^{2}+b^{2}}\geq b ,
\end{equation}
in terms of which the lapse function is exactly the RN-AdS one,
\begin{equation}\label{eq:FR}
f(r)=F(R),\qquad F(R)=1-\frac{2M}{R}+\frac{Q^{2}}{R^{2}}+\frac{R^{2}}{l^{2}} .
\end{equation}
All the $b$-dependence of the spacetime therefore sits in the relation between $r$ and $R$, i.e., in the radial part of the metric, $dr^{2}=R^{2}dR^{2}/(R^{2}-b^{2})$. This observation simplifies much of what follows.
\begin{figure}
\includegraphics[width=0.75\linewidth]{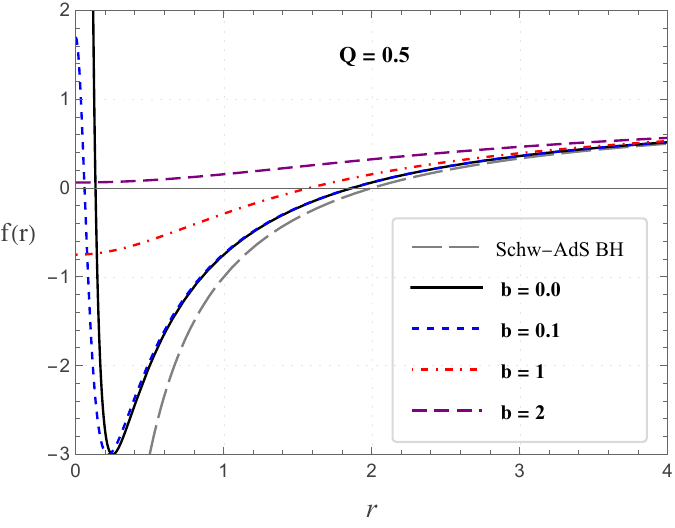}
\includegraphics[width=0.75\linewidth]{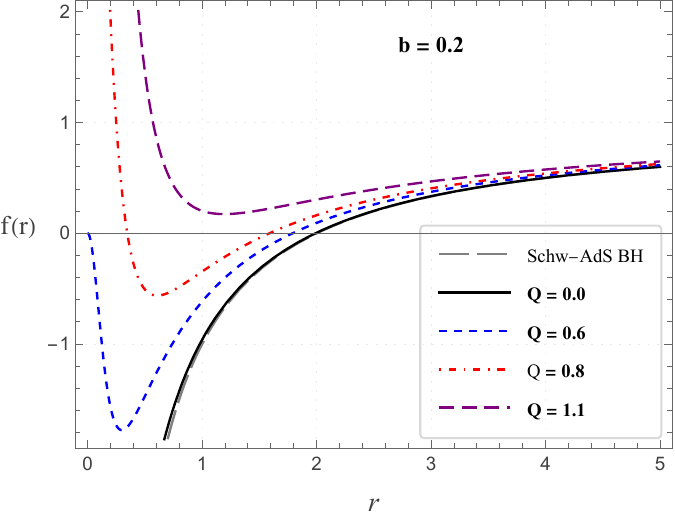}
\caption{Radial profile of the metric function $f(r)$ at $l=500M$.}\label{Fig:function_f}
\end{figure}
\begin{figure*}[ht!]
\centering
\includegraphics[width=0.3\linewidth]{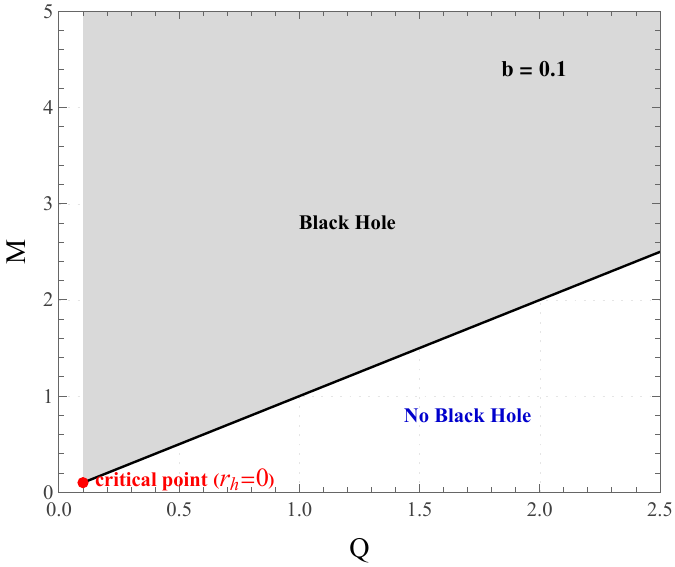}
\includegraphics[width=0.3\linewidth]{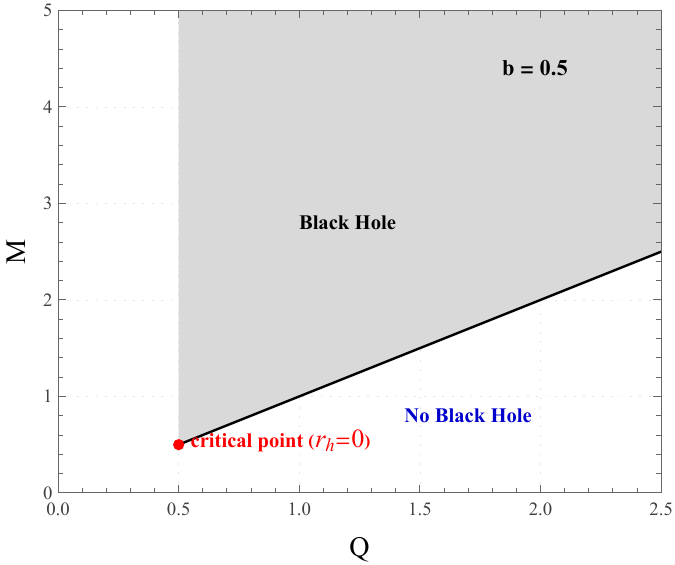}
\includegraphics[width=0.3\linewidth]{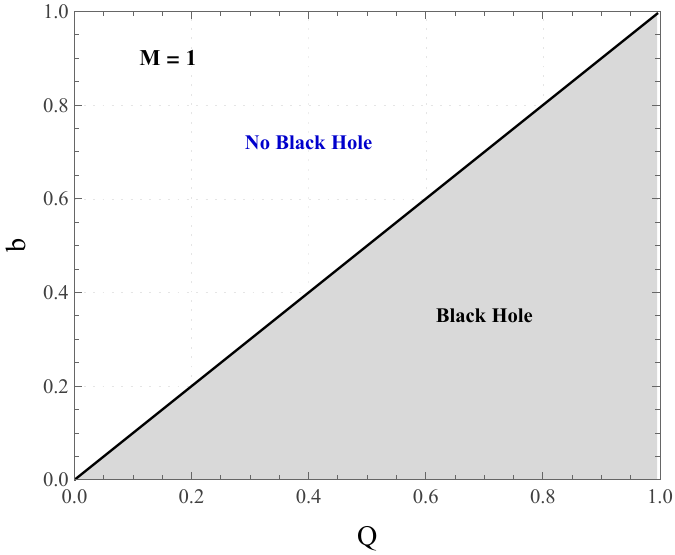}
\caption{Regions of parameter space with and without a horizon, in the $M$--$Q$ and $b$--$Q$ planes, at $l=500M$.}\label{fig:BHregion}
\end{figure*}
\begin{figure*}[ht!]
\centering
\includegraphics[width=0.3\linewidth]{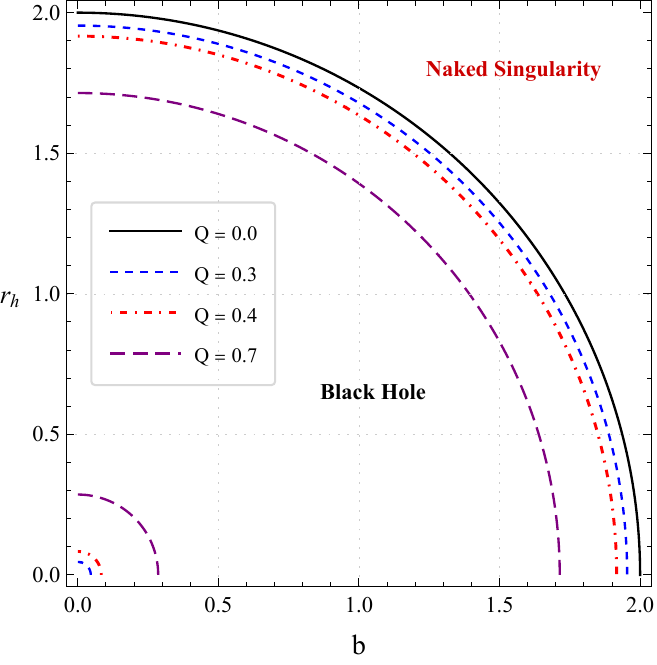}
\includegraphics[width=0.3\linewidth]{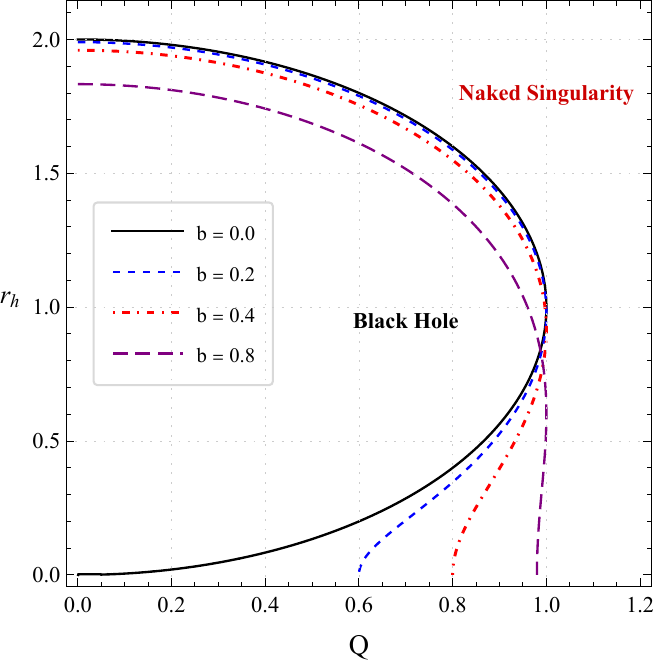}
\includegraphics[width=0.3\linewidth]{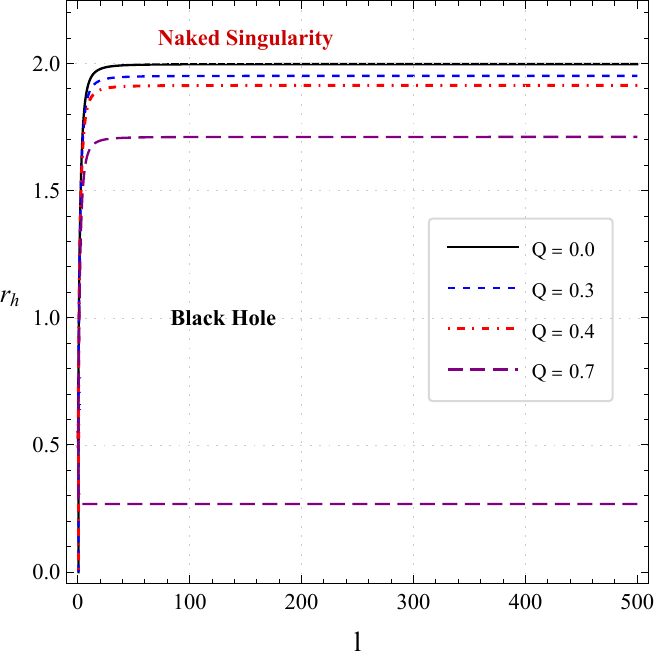}
\caption{Inner and outer horizons of the charged Simpson--Visser--AdS spacetime as functions of $b$ (left), $Q$ (middle) and $l$ (right). In the left and right panels $l=500M$, while in the middle panel $b=0.1M$.}\label{fig:horizons}
\end{figure*}

Figure~\ref{Fig:function_f} displays the radial profile of the metric function $f(r)$, which determines the causal structure. At a given $r$, the RN-AdS lapse lies above the Schwarzschild--AdS one because of the positive $Q^{2}$ term, and the charged Simpson--Visser--AdS lapse lies higher still in the strong-field region, since $f(r)=F(R)$ is evaluated at $R>r$ where $F$ is larger.

The horizons are the roots $r_{\pm}=\sqrt{R_{\pm}^{2}-b^{2}}$, where $R_{\pm}$ are the inner and outer roots of $F(R)=0$, i.e., the positive roots of $R^{4}+l^{2}R^{2}-2Ml^{2}R+l^{2}Q^{2}=0$. A horizon exists only if the corresponding $R_{\pm}$ exceeds $b$. This gives the following classes of geometry:
\begin{enumerate}
\item[I)] $b<R_{-}$: a regular black hole with an inner (Cauchy) and an outer (event) horizon on each side of the throat;
\item[II)] $R_{-}<b<R_{+}$ (this includes every $b<R_{+}$ when $Q=0$): a black bounce with a single horizon; the throat at $r=0$ lies inside the trapped region;
\item[III)] $b=R_{+}$: an extremal null throat (a one-way wormhole);
\item[IV)] $b>R_{+}$, or $Q>Q_{\rm ext}$ with $b>0$: a horizonless, two-way traversable wormhole with $f(r)>0$ everywhere.
\end{enumerate}
Here $Q_{\rm ext}(l)$ is the charge at which $R_{-}=R_{+}$; it satisfies $Q_{\rm ext}^{2}=MR_{\rm ext}+R_{\rm ext}^{4}/l^{2}$, with $R_{\rm ext}(1+2R_{\rm ext}^{2}/l^{2})=M$. For instance, $Q_{\rm ext}\simeq0.99995M$ at $l=100M$, and $Q_{\rm ext}\to M$ as $l\to\infty$. For $b=0$ and $Q>Q_{\rm ext}$ the RN-AdS naked singularity is recovered, whereas any $b>0$ removes it.

Both the electric charge $Q$ and the regularization parameter $b$ reduce the outer horizon radius $r_{+}$ compared with the uncharged, singular black hole. Once the parameters cross the thresholds above (e.g., $Q=1.1M$ with $b=0.2M$, or $Q=0.5M$ with $b=2M$ in Fig.~\ref{Fig:function_f}), $f(r)>0$ for all $r$ and the horizons disappear. Because $b\neq0$, this regime describes a regular, horizonless compact object rather than a naked singularity.

Figure~\ref{fig:BHregion} shows the corresponding phase diagram, separating geometries with an event horizon from horizonless ones. For $b\neq0$ a horizon exists only if $R_{+}(M,Q,l)>b$. In the $M$--$Q$ plane the black hole region therefore no longer reaches the origin: for $Q=0$ it starts at $M\simeq b/2$ rather than at $M=0$, and the boundary $R_{+}=b$ (where $r_{+}=0$) is shifted to larger $M$ as $b$ grows. Since $Q$ and $b$ both lower $r_{+}$, together they shrink the region where black holes exist, and a horizonless configuration is reached at smaller values of each parameter.

Figure~\ref{fig:horizons} shows the inner and outer horizons as functions of $b$ (left panel), $Q$ (middle panel), and $l$ (right panel). As $l$ increases, the outer horizon $r_+$ grows at first and then approaches its asymptotically flat value; a larger $l$ means a weaker negative cosmological constant. As $Q$ increases, $r_+$ decreases while the inner horizon $r_-$ moves outward, until the two merge at $Q=Q_{\rm ext}$, which does not depend on $b$ as long as $R_{\rm ext}>b$. At fixed $(Q,l)$ both $r_{\pm}=\sqrt{R_{\pm}^{2}-b^{2}}$ decrease monotonically with $b$, and the inner horizon exists only if $R_{-}>b$. For $b=0.1M$ (middle panel) this requires $Q\gtrsim0.44M$, so the inner-horizon branch starts at this value of the charge.
\section{Circular motion around charged Simpson-Visser-AdS black holes}\label{sec3: circular motion}
This section examines the motion of a neutral test particle of mass $m$, with $p^{\mu}p_{\mu}=-m^2$, in the static, spherically symmetric charged Simpson--Visser--AdS spacetime. Henceforth, the corresponding equation of motion for metric {given in Eq.~\eqref{metric}, can be obtained by using the Lagrangian density~\cite{2022MPLA...3750064K} as,
\begin{equation}\label{lagrangian1}
\mathscr{L}_p=\frac{1}{2}m\, g_{\mu \nu} {u}^{\mu} {u}^\nu .
\end{equation}
The corresponding time translation and rotational symmetry of the spacetime geometry related to conserved quantities can be determined through the Killing vectors
\begin{equation}
\xi_{(t)}^{\mu}\partial_{\mu}=\partial_{t} , \qquad
\xi_{(\phi)}^{\mu}\partial_{\mu}=\partial_{\phi},
\end{equation}
with $\xi_{(t)}^{\mu}=(1,~0,~0,~0)$ and $\xi_{(\phi)}^{\mu}=(0,~0,~0,~1)$. The associated conserved quantities are the specific energy $\mathcal{E}=E/m$ and the specific angular momentum $\mathcal{L}=L/m$ of the particle. In terms of the metric coefficients, they take the form
\begin{equation}\label{Energymom}
\mathcal{E}=-\frac{p_{t}}{m}=f(r)\dot{t}, \qquad
\mathcal{L}=\frac{ p_{\phi}}{m} = (r^2+b^2)\sin^2\theta\, \dot{\phi},
\end{equation}
where the dot denotes the derivative with respect to the proper time $\tau$. The expressions in Eq. \eqref{Energymom}, can equivalently be expressed as,
\begin{equation}
\dot{t}=\frac{\mathcal{E}}{f(r)},\qquad \dot{\phi}=\frac{\mathcal{L}}{(r^2+b^2) \sin^2\theta}.
\end{equation}
The normalization of the four-velocity reads
\begin{equation}\label{norm4vel}
g_{\mu \nu}u^{\mu}u^{\nu}=\epsilon \ .
\end{equation}
In the above expression, $\epsilon$ takes the values $-1$ and $0$ for massive and massless particles, respectively. The corresponding equation of motion for a particle with non-zero mass is governed by timelike geodesics. Henceforth, the equations of motion can easily be calculated utilizing the condition in Eq.~\eqref{norm4vel},
\begin{eqnarray}\label{eqmotionneutral}
\dot{r}^2&=&{\cal E}^2 - f(r)\left(1+\frac{{\cal L}^2}{r^2+b^2}\right)\ ,
 \\
\dot{\theta}^2&=&\frac{1}{g_{\theta \theta}^2}\Big({\cal K}-\frac{{\cal L}^2}{\sin^2\theta}\Big)\ ,
\end{eqnarray}
where ${\cal K}$ is the Carter constant. Because of spherical symmetry we can restrict the motion to the equatorial plane, $\theta=\pi/2$ and $\dot{\theta}=0$. On making use of the normalization condition $g_{\mu\nu}\dot{x}^\mu \dot{x}^\nu = -1$, the radial equation of motion takes the form
\begin{eqnarray}
 \dot{r}^2={\cal E}^2-V_{\rm eff}\ ,
\end{eqnarray}
with the effective potential
\begin{eqnarray}\label{effpotentail}
V_{\text{eff}} = f(r)\left(1+\frac{{\cal L}^2}{r^2+b^2}\right)=F(R)\left(1+\frac{{\cal L}^2}{R^2}\right) .
\end{eqnarray}
The second form shows that, at fixed $\mathcal{L}$, the effective potential is the RN-AdS potential written as a function of $R$; the regularization only relabels the radial position.
\begin{figure*}[ht!]
\centering
  \includegraphics[width=0.325\linewidth]{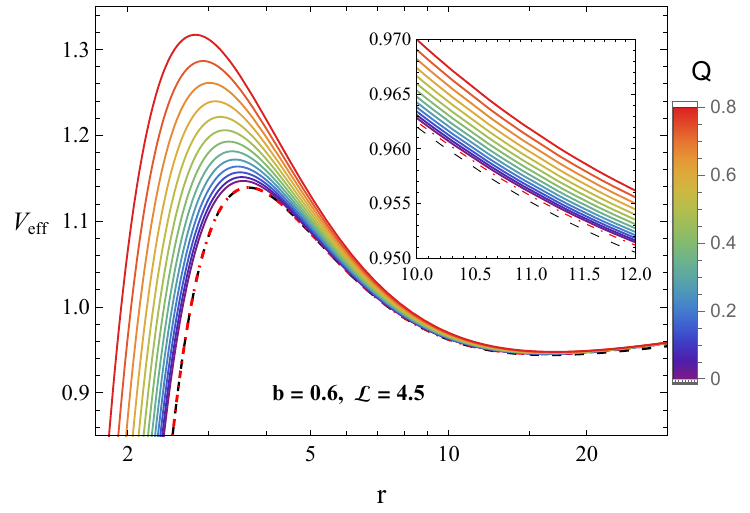}
  \includegraphics[width=0.325\linewidth]{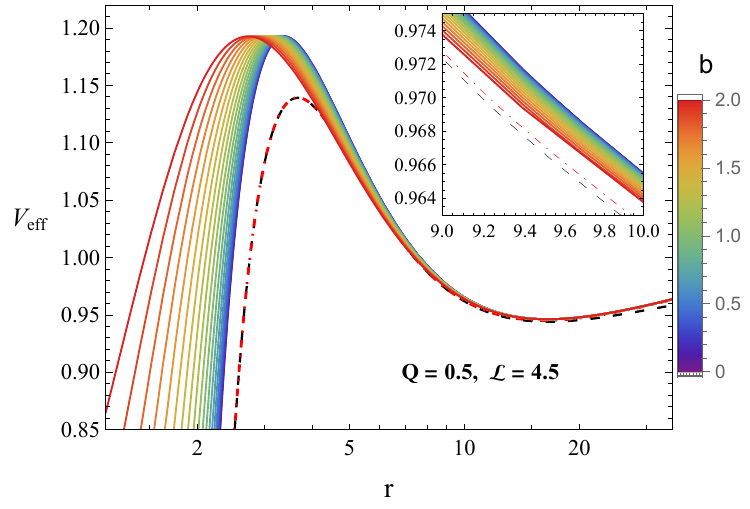}
     \includegraphics[width=0.325\linewidth]{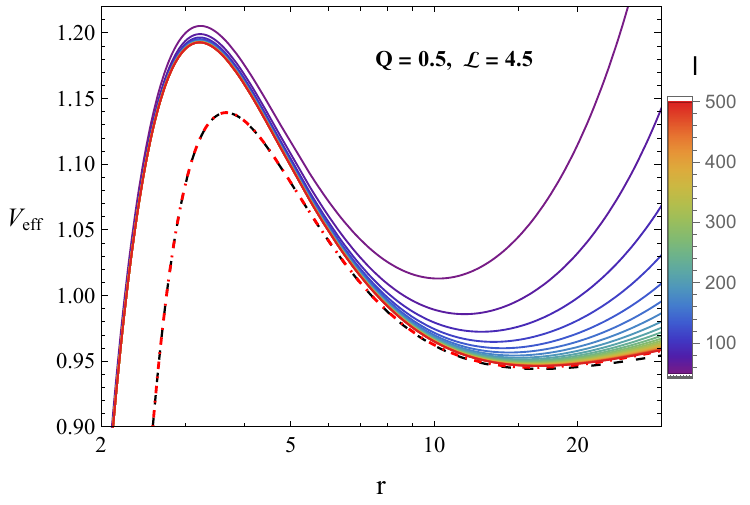}
   \caption{Effective potential of a test particle at $l=500M$ (left and middle panels) and $b=1M$ (right panel). The black dashed and red dot-dashed curves correspond to the Schwarzschild and Schwarzschild--AdS black holes, respectively.}\label{plotVeff}
\end{figure*}

In Fig.~\ref{plotVeff} we plot the effective potential for several combinations of $Q$, $b$, and $l$ and compare it with the Schwarzschild (black dashed) and Schwarzschild--AdS (red dot-dashed) cases. The extrema, $\partial_r V_{\text{eff}} = 0$, give the circular orbits: minima correspond to stable and maxima to unstable orbits.

The charge adds a positive term $Q^{2}(1+\mathcal{L}^{2}/R^{2})/R^{2}$ to the potential. This raises the barrier at small radii, i.e., the maximum associated with unstable circular orbits, and at fixed $\mathcal{L}$ it moves the stable minimum outward. The AdS term contributes $(R^{2}+\mathcal{L}^{2})/l^{2}$, which grows with $R$, so that $V_{\rm eff}\to\infty$ as $R\to\infty$: in an AdS background every massive particle is bound, and none can escape to infinity. Increasing $l$ weakens this confinement, and for $l=500M$ the potential is already very close to the Schwarzschild one over the plotted range. Finally, the extremal values of $V_{\rm eff}$ do not depend on $b$; in the $r$ coordinate, $b$ only shifts the curves toward $r=0$, by $r=\sqrt{R^{2}-b^{2}}$.
\subsection{Circular orbits}\label{subsec: circular orbits}
To determine circular geodesics, one can use the following two simultaneous conditions
\beq
V_{\rm eff}(r) = \mathcal{E}^2, \quad \frac{\partial V_{\rm eff}(r)}{\partial r} = 0,
\eeq
which give the specific angular momentum $\mathcal{L}$ and energy $\mathcal{E}$ for circular orbits, and can be derived as
\begin{equation}
  \mathcal{E}^{2}=\frac{2r f(r)^{2}}{\mathcal{D}},\qquad
  \mathcal{L}^{2}=\frac{(r^2+b^2)^2f^{\prime}(r)}{\mathcal{D}},
  \label{eq:EL}
\end{equation}
with
\beq
\mathcal{D}\equiv 2rf(r)-(r^2+b^2)f^{\prime}(r) ,
\eeq
where a prime denotes the derivative with respect to $r$. Since $f^{\prime}(r)=(r/R)F^{\prime}(R)$, the factors of $r$ cancel and
\begin{equation}\label{eq:EL_R}
\mathcal{E}^{2}=\frac{2F^{2}}{2F-RF^{\prime}},\qquad
\mathcal{L}^{2}=\frac{R^{3}F^{\prime}}{2F-RF^{\prime}},
\end{equation}
where now the prime acts on $R$. These are the RN-AdS expressions evaluated at the areal radius, so at fixed $R$ the energy and angular momentum of a circular orbit are independent of $b$.
%
\subsection{Angular momentum and energy}\label{subsec: angular momentum}
In this subsection, we examine the angular momentum and energy of a particle orbiting a regular charged Simpson-Visser-AdS black hole.

\begin{figure*}[ht!]
    \includegraphics[width=0.3\textwidth]{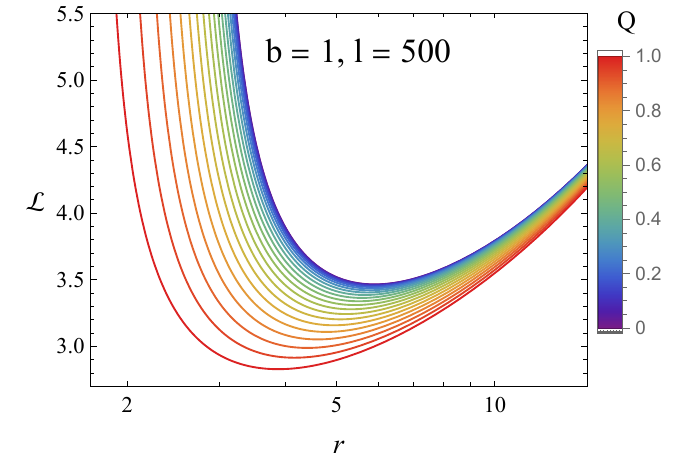}
    \includegraphics[width=0.3\textwidth]{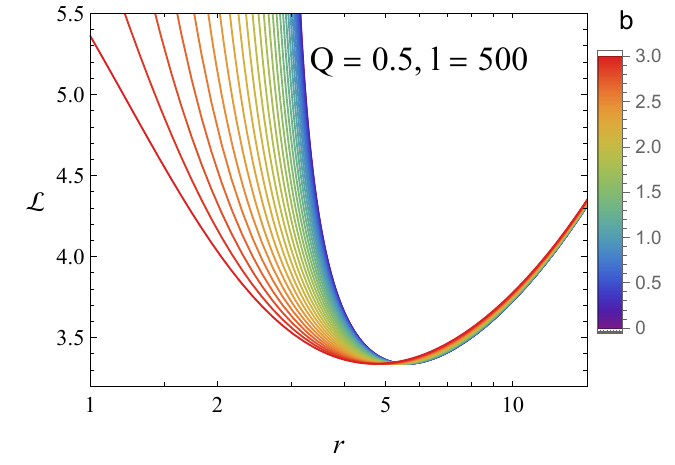}
    \includegraphics[width=0.3\textwidth]{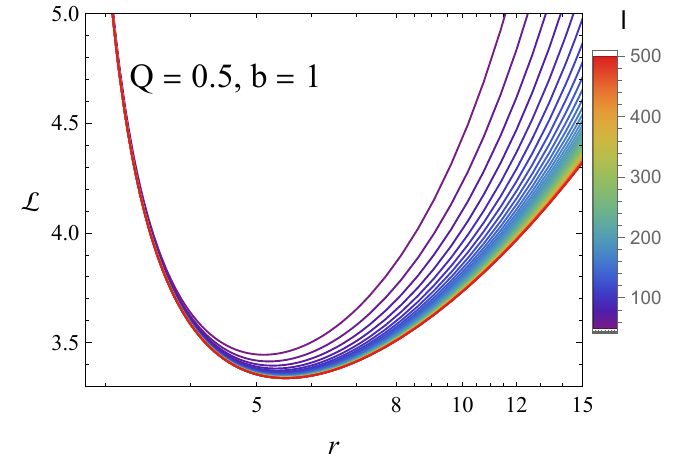}
    \includegraphics[width=0.3\textwidth]{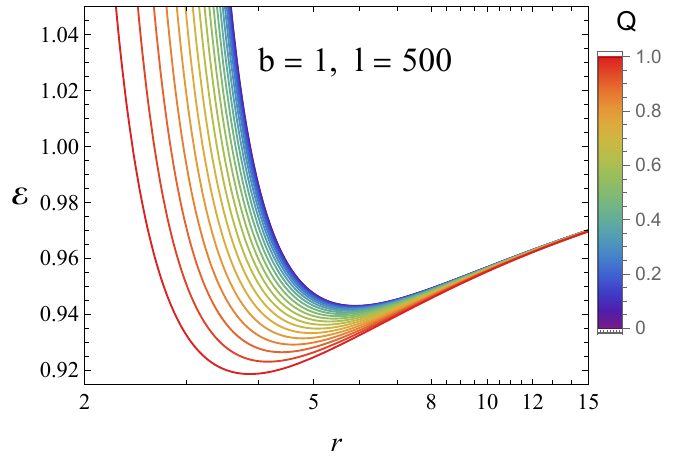}
    \includegraphics[width=0.3\textwidth]{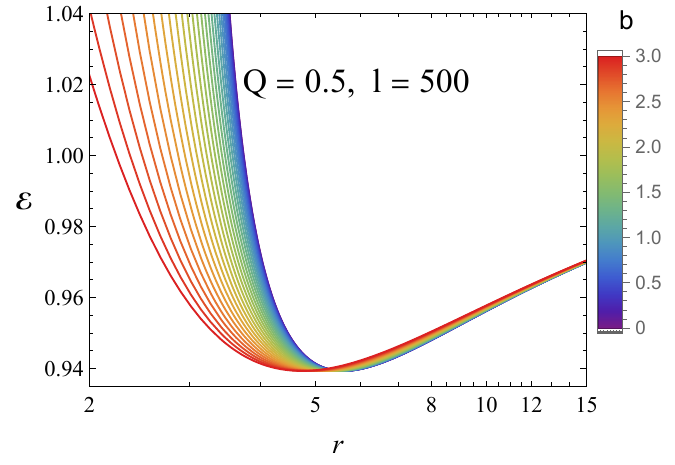}
    \includegraphics[width=0.3\textwidth]{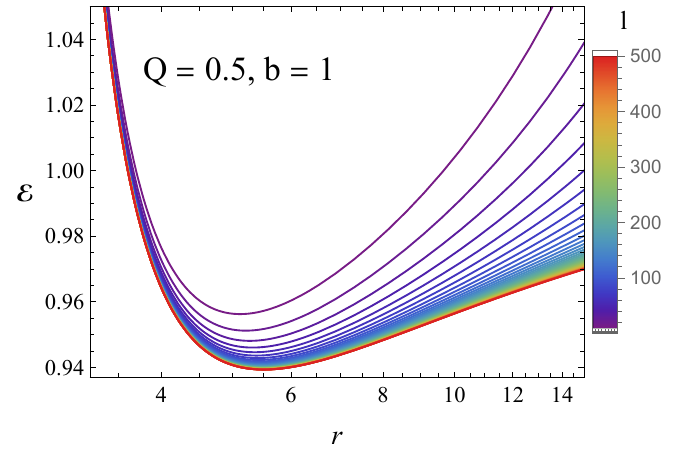}
\caption{Radial profiles of the specific angular momentum (top row) and specific energy (bottom row) of circular orbits for several values of the spacetime parameters.}\label{Fig:A.mom_eng}
\end{figure*}
Figure~\ref{Fig:A.mom_eng} shows the radial profiles of the specific angular momentum $\mathcal{L}$ and specific energy $\mathcal{E}$ of circular orbits. The minimum of each curve marks the ISCO. Increasing $Q$ moves the ISCO inward and lowers both minima; for example, $\mathcal{E}_{\rm ISCO}$ drops from $0.9428$ for Schwarzschild to $0.9392$ at $Q=0.5M$ and $0.9313$ at $Q=0.8M$ (Table~\ref{tab:char}). The AdS radius acts differently. A smaller $l$ moves the ISCO inward but raises $\mathcal{E}_{\rm ISCO}$ and $\mathcal{L}_{\rm ISCO}$, because the confining AdS term adds to the energy of every orbit. Increasing $l$ therefore lowers both minima and pushes the ISCO back toward its asymptotically flat value, and for $l=500M$ the curves are close to the flat-space ones.

The regularization parameter $b$ behaves differently from both. By Eq.~\eqref{eq:EL_R}, the minimum values $\mathcal{E}_{\rm ISCO}$ and $\mathcal{L}_{\rm ISCO}$ do not depend on $b$; in the $r$ coordinate the whole curve is merely shifted to smaller $r$. At a fixed coordinate radius $r$ outside the ISCO the orbit sits at a larger areal radius $R$, so the curves for larger $b$ lie slightly higher there. This shift is kinematic; it does not mean that $b$ weakens gravity or changes the binding energy at the ISCO.
\subsection{Innermost stable circular orbits}\label{subsec: ISCO}
The ISCO is the circular orbit at which stability is lost, i.e., where both $\partial_r V_{\rm eff}=0$ and $\partial_{r}^{2}V_{\rm eff}=0$. Equivalently, it is the outermost zero of the radial epicyclic frequency~\eqref{eq:omega2} derived below. In terms of $f(r)$ the condition reads
\begin{equation}\label{ISCO}
r f f^{\prime\prime}-2r\left(f^{\prime}\right)^{2}+\frac{3r^{2}-b^{2}}{r^{2}+b^{2}}\,f f^{\prime}=0 .
\end{equation}
Using Eq.~\eqref{eq:FR}, this reduces to a condition on $R$ alone, which for the lapse~\eqref{eq:FR} becomes the polynomial equation
\begin{eqnarray}
&&MR^{3}-6M^{2}R^{2}+9MQ^{2}R-4Q^{4} \nonumber\\
&&\quad+\frac{R^{4}}{l^{2}}\left(4R^{2}-15MR+12Q^{2}\right)=0 .\label{eq:isco_poly}
\end{eqnarray}
Its largest root outside the horizon, $R_{\rm ISCO}(Q,l)$, does not depend on $b$, and the ISCO coordinate radius is
\begin{equation}\label{eq:risco}
r_{\rm ISCO}=\sqrt{R_{\rm ISCO}^{2}-b^{2}} \qquad (b<R_{\rm ISCO}) .
\end{equation}
For $b\geq R_{\rm ISCO}$, which can only happen for wormholes, stable circular orbits extend down to the throat. Equation~\eqref{eq:isco_poly} gives $R_{\rm ISCO}=6M$ for Schwarzschild and reproduces the known RN result for $l\to\infty$.
%
\begin{figure*}[ht!]
\centering
    \includegraphics[width=0.28\linewidth]{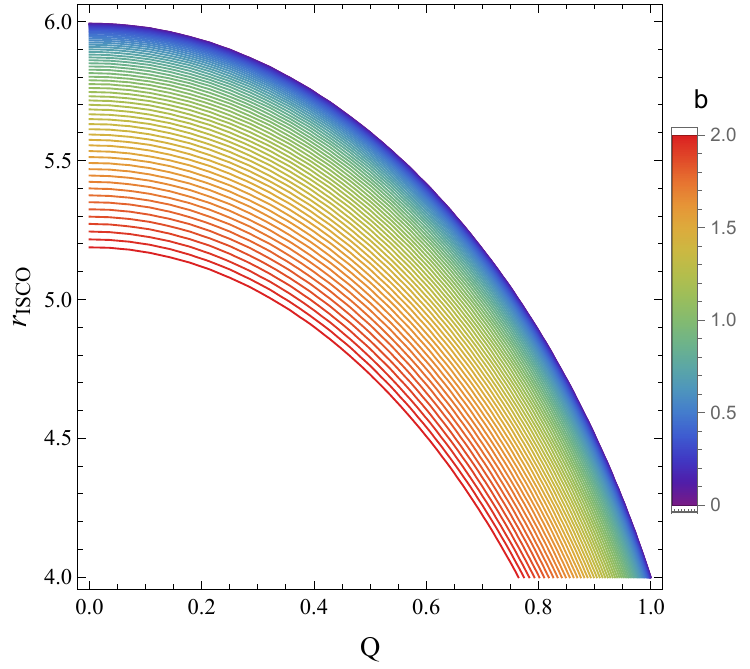}
    \includegraphics[width=0.28\linewidth]{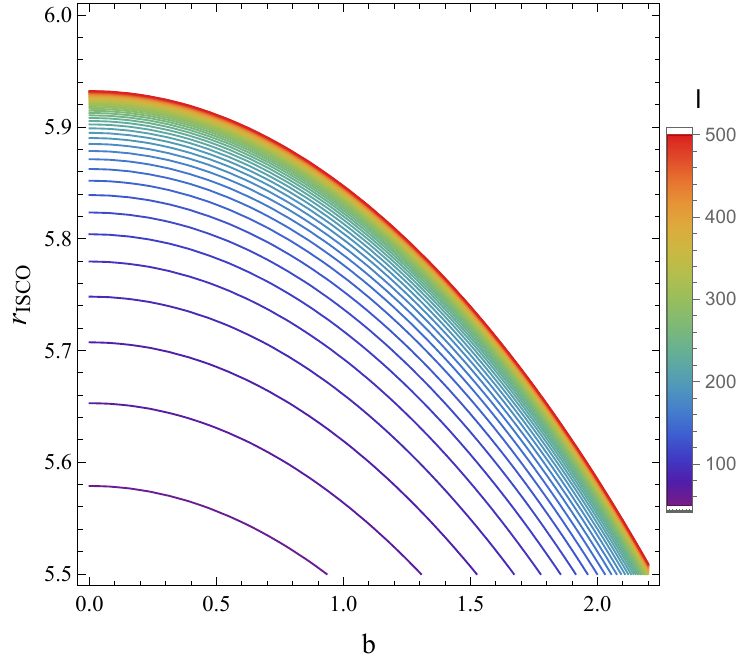}
    \includegraphics[width=0.28\linewidth]{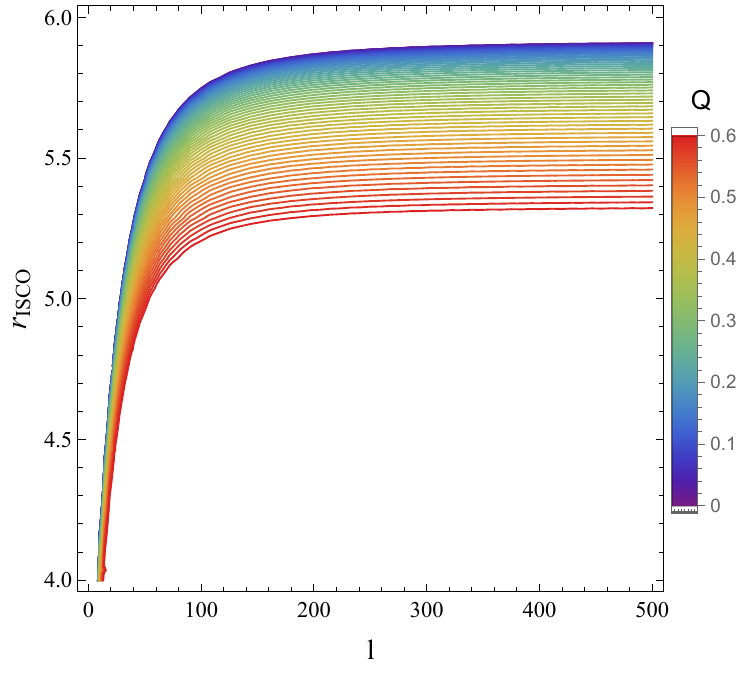}
\caption{ISCO radius $r_{\rm ISCO}$ as a function of the charge $Q$, the regularization parameter $b$, and the AdS radius $l$, at $l=500M$ (left), $Q=0.2M$ (middle), and $b=1M$ (right).\label{fig:ISCO}}
  \end{figure*}
%
Figure~\ref{fig:ISCO} shows how $r_{\rm ISCO}$ depends on $b$, $l$, and $Q$. It decreases monotonically with both $Q$ and $b$, but for different reasons. The charge genuinely moves the ISCO inward: $R_{\rm ISCO}$ drops from $6M$ to $5.607M$ at $Q=0.5M$ and to $4.891M$ at $Q=0.8M$ (Table~\ref{tab:char}). The regularization parameter only changes the coordinate radius through Eq.~\eqref{eq:risco}; for $Q=0.5M$ and $l=100M$, for example, $r_{\rm ISCO}$ goes from $5.475M$ to $5.383M$, $5.097M$ and $4.580M$ for $b=0,1,2,3M$, while $R_{\rm ISCO}$, $\mathcal{E}_{\rm ISCO}$, $\mathcal{L}_{\rm ISCO}$ and the orbital frequency at the ISCO stay the same.

The dependence on the AdS radius is strongly nonlinear. For small $l$ the AdS term in Eq.~\eqref{eq:isco_poly} is important and the ISCO moves well inside $6M$, e.g., $R_{\rm ISCO}=5.834M$ at $l=100M$ and $5.372M$ at $l=40M$ for $Q=b=0$. As $l$ increases the spacetime approaches asymptotic flatness ($\Lambda \to 0$), the ISCO moves outward, and it converges to the value of the asymptotically flat charged Simpson--Visser black hole; at $l=1000M$ the difference is already only $2\times10^{-3}M$.
\section{Fundamental frequencies and QPOs}\label{sec4:Fund_frequencies and QPOs}
This section deals with the fundamental frequencies of test-particle motion near a charged Simpson--Visser--AdS black hole. Small perturbations of stable circular orbits are described by the orbital (Keplerian) frequency together with the radial and vertical epicyclic frequencies. These frequencies characterize the motion of matter in the strong gravitational field close to the black hole, and they are the basic ingredients of the models used to interpret twin-peak QPOs. Their behavior therefore tells us which values of the upper and lower QPO frequencies a given spacetime can produce.
\subsection{Harmonic oscillations}\label{subsec: Harmonic_Oscs}
Test particles on stable circular orbits in the equatorial plane undergo small oscillations in the radial and vertical directions. We obtain the corresponding frequencies by perturbing a circular orbit, $r\to r_0+\delta r$ and $\theta\to \theta_0+\delta\theta$ with $\theta_0=\pi/2$.

Expanding the effective potential around $x_0= (r_0, \theta_{0})$, we obtain
\begin{eqnarray}\label{Vexpand}
\nonumber
V_{\rm eff}(r, \theta) &=& V_{\rm eff}(r_0,\theta_0) +\delta r\,\partial_r V_{\rm eff}(r,\theta)\Big|_{x_0}
\\\nonumber
&& +\delta\theta\, \partial_\theta V_{\rm eff}(r,\theta)\Big|_{x_0} +\frac{1}{2}\delta r^2\,\partial_r^2 V_{\rm eff}(r,\theta)\Big|_{x_0}
\\\nonumber
&& +\frac{1}{2}\delta\theta^2\,\partial_\theta^2 V_{\rm eff}(r,\theta)\Big|_{x_0} +\delta r\,\delta\theta\,\partial_r\partial_\theta V_{\rm eff}(r,\theta)\Big|_{x_0}\\&&
+{\cal O}\left(\delta r^3,\delta\theta^3\right)\, .
\end{eqnarray}
The zeroth-order term equals $\mathcal{E}^{2}$ and cancels in the equation of motion $\dot r^{2}=\mathcal{E}^{2}-V_{\rm eff}$. The first-order terms vanish because a circular orbit is an extremum of $V_{\rm eff}$, and the mixed term vanishes by reflection symmetry about the equatorial plane. Only the second derivatives remain, so $\delta r$ and $\delta\theta$ oscillate independently. Converting from proper time to the Killing time $t$ with $u^{t}=dt/d\tau=\mathcal{E}/f$, we obtain the frequencies measured with respect to $t$:
\bea
&& \Omega_\phi = \sqrt{\frac{f'(r)}{2r}}, \label{eq:Omphi}\\
&&\Omega_r = \sqrt{\frac{f(r)^2}{2\mathcal{E}^2} \frac{\partial^2 V_{\text{eff}}}{\partial r^2}}\bigg|_{\theta=\pi/2},\label{eq:Omr_def}\\
&&\Omega_\theta = \sqrt{\frac{f(r)}{2\mathcal{E}^2\,(r^{2}+b^{2})} \frac{\partial^2 V_{\text{eff}}}{\partial \theta^2}}\bigg|_{\theta=\pi/2}.\label{eq:Omth_def}
\eea
DThe straightforward differentiation of the effective potential results in the mathematical equations for the radial frequency profile
\bea \nonumber
\Omega_r^2 &= &\frac{f^2(r)}{2 \EE^2} \bigg[ f^{\prime\prime}(r)\left(1+\frac{ \LL^2}{r^2+b^2}\right)  \\
            && - \frac{4r \LL^2 f^{\prime}(r)}{(r^2+b^2)^2}- \frac{2 \LL^2 f(r)(b^2-3r^2)}{(r^2+b^2)^3} \bigg]. \label{eq:omega1}
\eea
Eliminating $\mathcal{E}^{2}$ and $\mathcal{L}^{2}$ with Eq.~\eqref{eq:EL}, Eq.~\eqref{eq:omega1} takes the closed form
\begin{equation}
  \Omega_{r}^{2}=\tfrac{1}{2}f(r) f^{\prime\prime}(r) -(f^{\prime}(r))^2
  +\frac{3r^{2}-b^{2}}{2r\bigl(r^{2}+b^{2}\bigr)}f(r) f^{\prime}(r).  \label{eq:omega2}
\end{equation}
Since the spacetime is spherically symmetric, every equatorial circular orbit can be rotated into an inclined one, and therefore $\Omega_{\theta}=\Omega_{\phi}$. The geometry cannot produce nodal (Lense--Thirring) precession, and we confront it only with the two high-frequency peaks.

With Eq.~\eqref{eq:FR}, both frequencies take a simple form in terms of the areal radius:
\begin{eqnarray}
\Omega_{\phi}^{2}&=&\frac{M}{R^{3}}-\frac{Q^{2}}{R^{4}}+\frac{1}{l^{2}}, \label{eq:Omphi_R}\\
\Omega_{r}^{2}&=&\left(1-\frac{b^{2}}{R^{2}}\right)\bigg[\frac{MR^{3}-6M^{2}R^{2}+9MQ^{2}R-4Q^{4}}{R^{6}} \nonumber\\
&&+\frac{1}{l^{2}}\left(4-\frac{15M}{R}+\frac{12Q^{2}}{R^{2}}\right)\bigg]. \label{eq:Omr_R}
\end{eqnarray}
The bracket in Eq.~\eqref{eq:Omr_R} is the radial epicyclic frequency of the RN-AdS black hole at radius $R$. Two consequences follow. First, at fixed $R$ the Keplerian frequency does not depend on $b$, and the regularization lowers only the radial frequency, by the factor $(1-b^{2}/R^{2})^{1/2}$. Second, the effect of $b$ dies off as $b^{2}/R^{2}$ far from the throat. For $b=Q=0$ and $l\to\infty$, Eq.~\eqref{eq:Omr_R} reduces to the Schwarzschild result $\Omega_r^{2}=M(R-6M)/R^{4}$; we used this limit to validate our numerical code. In an AdS background the frequencies do not vanish at large radii: $\Omega_{\phi}\to1/l$ and $\Omega_{r}\to2/l$ for $R\gg M$. As a result, $\Omega_{r}$ overtakes $\Omega_{\phi}$ at $R\simeq9.2M$, $13.4M$ and $27.9M$ for $l=40M$, $100M$ and $500M$ (with $Q=b=0$). Beyond these radii the relativistic-precession prescription introduced below would give $\nu_{L}<0$, so only orbits inside them are relevant for twin-peak QPOs.
\begin{figure}
\centering
\includegraphics[width=0.7\linewidth]{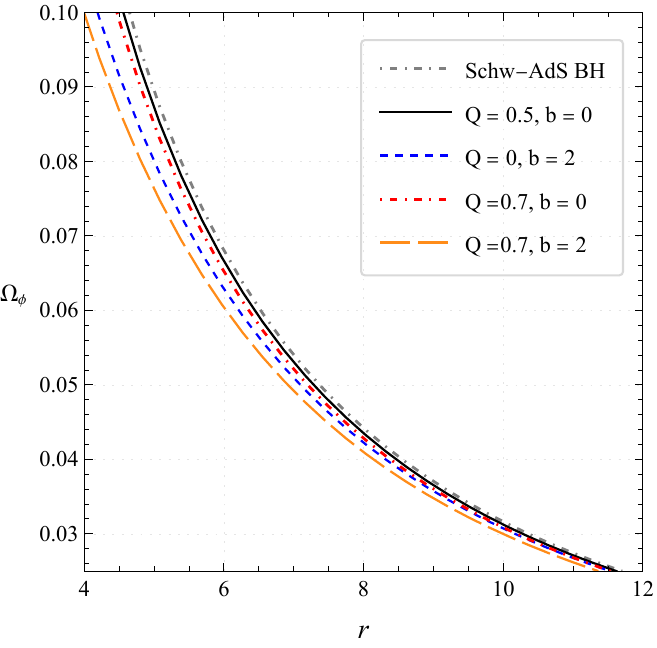}
\includegraphics[width=0.7\linewidth]{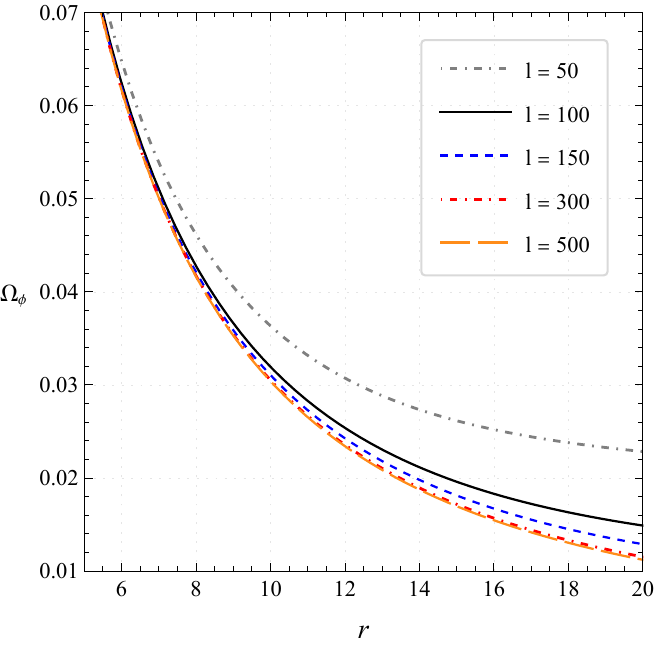}
\caption{Radial dependence of the Keplerian frequency of test particles in the charged Simpson--Visser--AdS spacetime. The top panel is plotted at $l=500M$; the bottom panel at $Q=0.5M$ and $b=2M$ (a horizonless wormhole configuration, since $b>R_{+}\simeq1.87M$).} \label{Fig:Kpfrequencies}
\end{figure}
\begin{figure}
\centering
\includegraphics[width=0.7\linewidth]{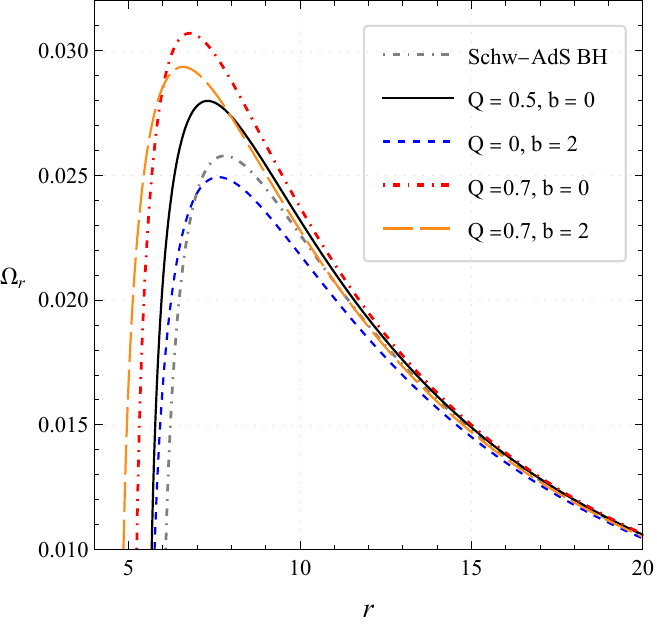}
\includegraphics[width=0.7\linewidth]{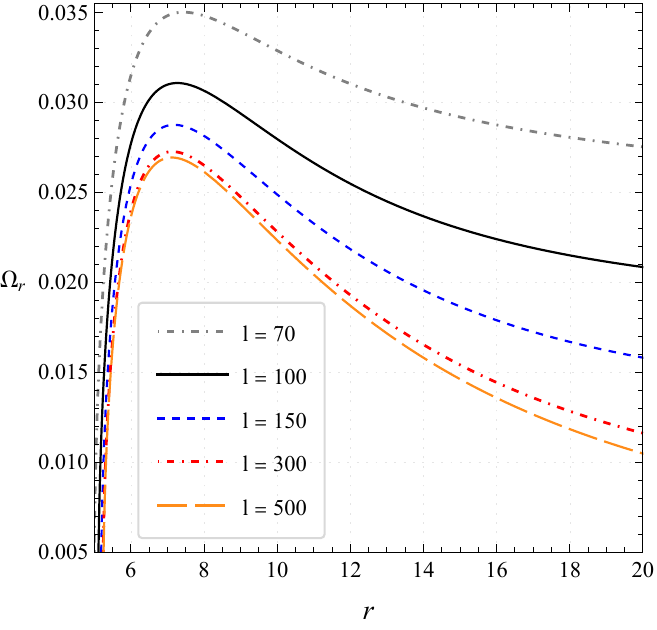}
\caption{Radial dependence of the radial epicyclic frequency of test particles in the charged Simpson--Visser--AdS spacetime. The top panel is plotted at $l=500M$; the bottom panel at $Q=0.5M$ and $b=2M$.} \label{Fig:Rdfrequencies}
\end{figure}
Figures~\ref{Fig:Kpfrequencies} and~\ref{Fig:Rdfrequencies} show the radial profiles of the Keplerian ($\Omega_\phi$) and radial epicyclic ($\Omega_r$) frequencies for different values of $Q$, $b$, and $l$. The Keplerian frequency decreases monotonically with radius, as in previous studies~\cite{Khan2026EPJC...86..597K}; in AdS it tends to $1/l$ instead of zero. The charge lowers $\Omega_{\phi}$ at fixed radius through the $-Q^{2}/R^{4}$ term in Eq.~\eqref{eq:Omphi_R}. The regularization parameter also lowers $\Omega_{\phi}$ at fixed $r$, but only because the orbit is then located at a larger areal radius $R$. The AdS radius acts in the opposite direction: the $1/l^{2}$ term raises $\Omega_{\phi}$, and its relative importance grows with distance, where it controls the deviation from asymptotic flatness.

For the radial epicyclic frequency the roles change. Increasing $b$ or $l$ lowers $\Omega_r$, while increasing $Q$ raises it. In the asymptotically flat limit the curves have a clear maximum, located at $R=8M$ for Schwarzschild. The charge raises this maximum and moves it inward, to $R\simeq7.51M$ for $Q=0.5M$ and $R\simeq6.63M$ for $Q=0.8M$. The regularization parameter and a larger $l$ lower the maximum. For small $l$ the maximum disappears altogether: for $l=40M$, $\Omega_{r}$ grows monotonically outside the ISCO toward $2/l$. Accordingly, among the configurations shown, the uncharged Simpson--Visser--AdS spacetime has the smallest $\Omega_r$, below both the Schwarzschild--AdS and the charged Simpson--Visser--AdS cases.

To express the frequencies in Hertz we restore physical units,
\begin{equation}\label{nu_i}
\nu_{i} = \frac{1}{2\pi}\frac{c^3}{GM}\, \Omega_{i}
\simeq 32.31\,{\rm kHz}\left(\frac{M_{\odot}}{M}\right)(M\Omega_{i}) ,
\end{equation}
where $i \in \{r, \theta, \phi\}$, $M\Omega_i$ is the dimensionless frequency, and we use the nominal solar mass parameter $GM_{\odot}/c^{3}=4.9255\,\mu{\rm s}$. In an asymptotically AdS spacetime there is no static observer at infinity with $f\to1$; as usual, $\nu_i$ refers to the Killing time $t$, i.e., to static observers in the region $M\ll R\ll l$ where $f\simeq1$. This identification holds up to corrections of order $R^{2}/l^{2}$, which we return to in Sec.~\ref{sec:qpo}.
\begin{figure*}
\centering
\includegraphics[width=0.31\linewidth]{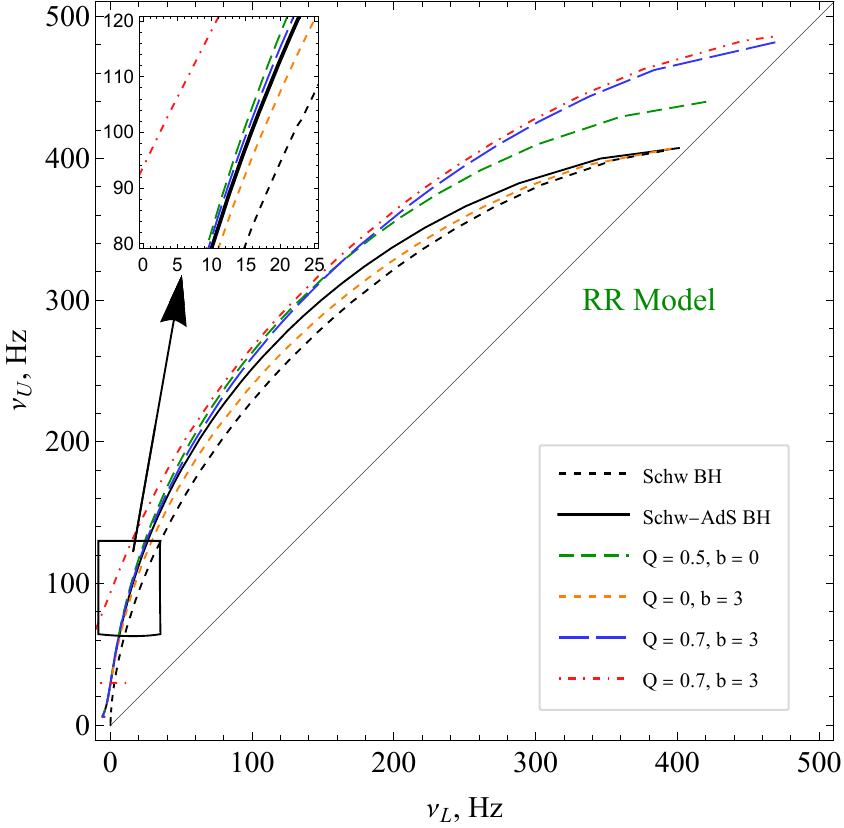}
\includegraphics[width=0.31\linewidth]{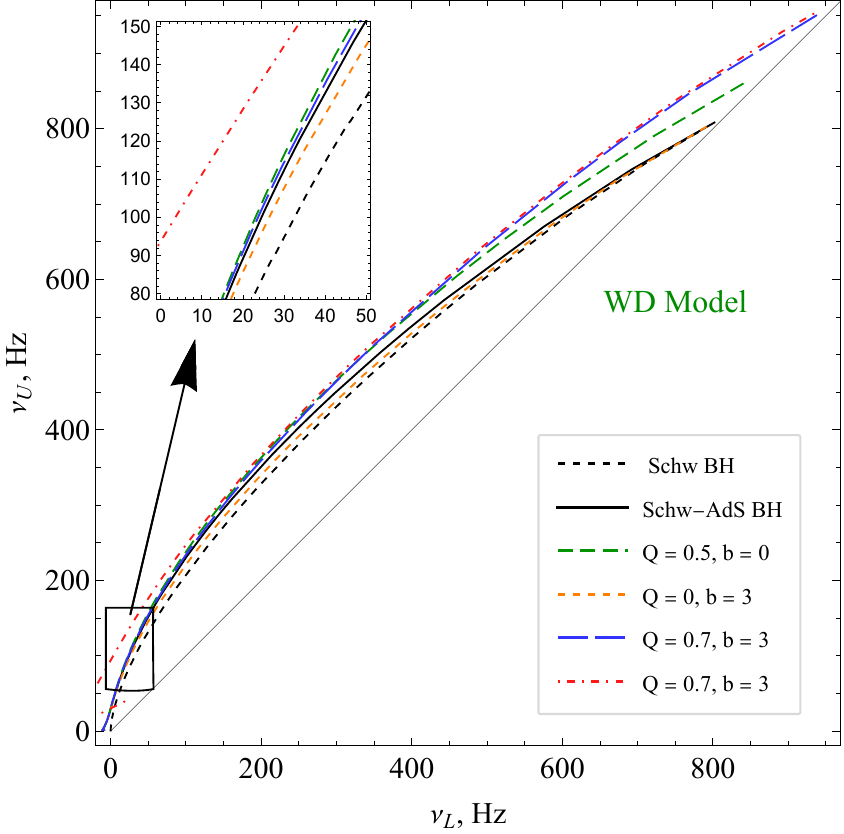}
\vspace{0.3cm}
\includegraphics[width=0.31\linewidth]{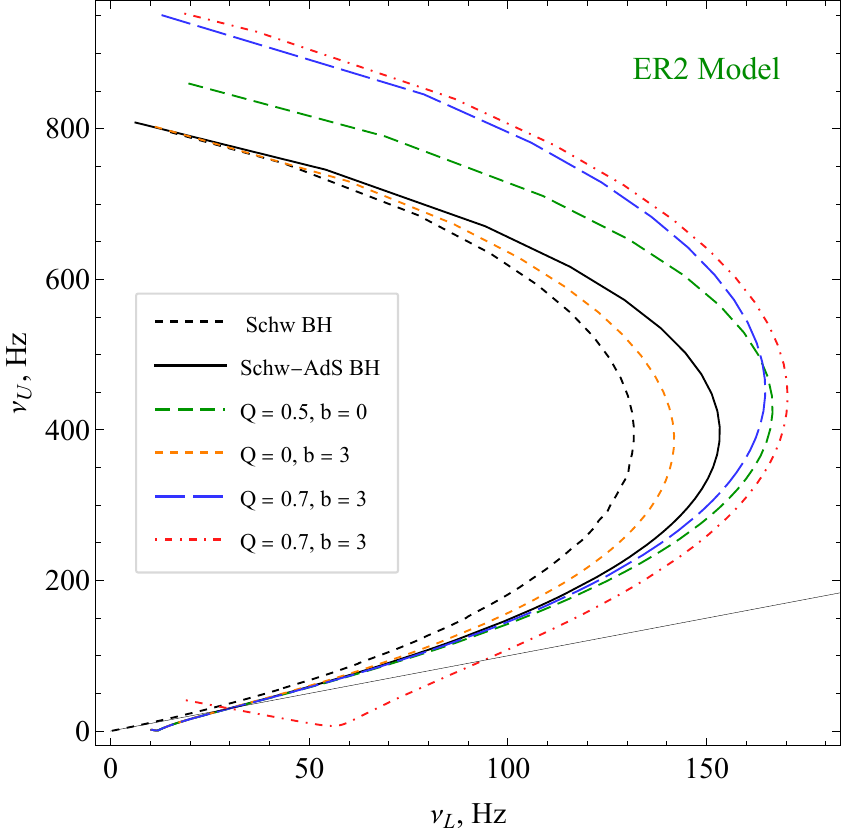}
\includegraphics[width=0.31\linewidth]{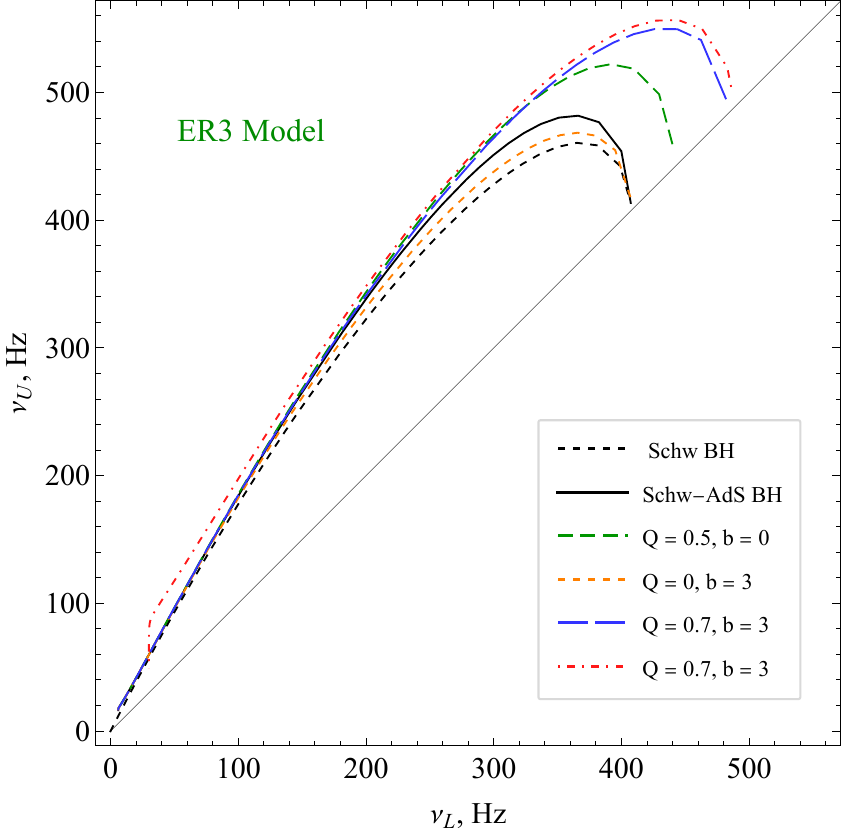}
\includegraphics[width=0.31\linewidth]{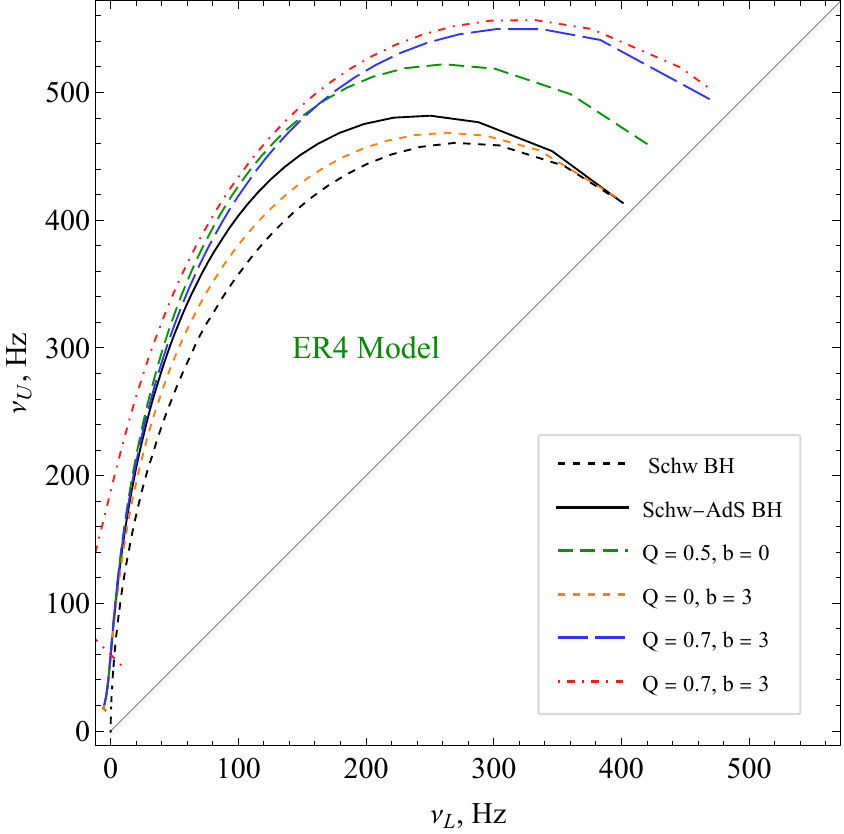}
\caption{Upper versus lower twin-peak QPO frequencies in the RP, WD, and ER2--ER4 models for the charged Simpson--Visser--AdS spacetime with $M=5.4M_{\odot}$, at $l=1000M$, except for the red dot-dashed curves, which are plotted at $l=200M$.}\label{QPOs}
\end{figure*}
\subsection{Quasi-periodic oscillations}\label{subsec:QPOs}
QPOs are among the most striking timing phenomena observed near compact objects, including black holes, neutron stars, and white dwarfs in binary systems~\cite{Ingram2010MNRAS,Zdunik2000AA}. They are seen most often in neutron-star binaries and less frequently in black-hole and white-dwarf systems. In the most common interpretation they arise from (quasi-)harmonic oscillations of accreting matter in the strong-field region, so their frequencies probe the spacetime geometry and can be used to constrain the properties of the central object. Applied to black holes, QPO data can therefore constrain the parameters of Simpson--Visser-type models and help distinguish between different black-hole metrics.

QPOs have also been used in this way for black holes in MOG~\cite{2024EPJC...84..964J,2025EPJC...85.1029N} and in exotic-field models, where frequency ratios and peak positions constrain both the surrounding matter and the underlying gravitational dynamics. They thus provide a route to testing phantom-field sources, such as those that support Simpson--Visser geometries, and alternative theories of gravity.

In this work we compute twin-peak QPO frequencies in the charged Simpson--Visser--AdS spacetime and compare them with the Schwarzschild and Schwarzschild--AdS cases. Following Ref.~\cite{2021EPJC...81.1067S}, we consider three classes of models. (I) The relativistic precession (RP) model~\cite{Stella1999ApJ}, in which the upper and lower frequencies are the azimuthal and periastron-precession frequencies, $\nu_U = \nu_\phi$ and $\nu_L = \nu_\phi - \nu_r$. (II) The warped disc (WD) model, which attributes the QPOs to oscillations in a thin, warped accretion disc, with $\nu_U = 2\nu_\phi - \nu_r$ and $\nu_L = 2(\nu_\phi - \nu_r)$. (III) The epicyclic resonance (ER) models, which assume a geometrically thick disc in which the QPOs arise from resonances between geodesic oscillation modes. We use three variants: ER2, $\nu_U = 2\nu_\theta - \nu_r$ and $\nu_L = \nu_r$; ER3, $\nu_U = \nu_\theta + \nu_r$ and $\nu_L = \nu_\theta$; and ER4, $\nu_U = \nu_\theta + \nu_r$ and $\nu_L = \nu_\theta - \nu_r$.

Figure~\ref{QPOs} shows the resulting tracks in the $\nu_U$--$\nu_L$ plane for a stellar-mass black hole with $M = 5.4M_\odot$ (the mass of GRO~J1655$-$40), compared with the Schwarzschild and Schwarzschild--AdS spacetimes. Each curve runs from the ISCO outward. Lines of small-integer frequency ratios (3:2, 4:3, 5:4) mark possible resonances, and the diagonal $\nu_U=\nu_L$ marks the limit where the two peaks merge into one; twin peaks require $\nu_U>\nu_L$. In the RP, WD, ER3 and ER4 models the tracks end on this diagonal at the ISCO, where $\nu_r=0$.

The three parameters shift the tracks in different ways (see Table~\ref{tab:char} for representative numbers). At a fixed lower frequency, a larger charge raises the upper frequency. In the RP model at $\nu_L=150$~Hz, for example, $\nu_U=281.5$~Hz for Schwarzschild, $292.7$~Hz for $Q=0.5M$ and $314.1$~Hz for $Q=0.8M$ (all at $l=1000M$, $b=0$). A smaller AdS radius works in the same direction ($\nu_U=287.7$~Hz for $Q=b=0$ and $l=200M$), and both $Q$ and a smaller $l$ also raise the orbital frequency at the ISCO, which in the RP model is the highest upper frequency a track can reach. The regularization parameter acts in the opposite direction and much more weakly. At fixed $\nu_U$, i.e., at fixed $R$, it raises $\nu_L$ by reducing $\nu_r$, so the $b\neq0$ tracks lie slightly below the $b=0$ track: at $\nu_L=150$~Hz and $Q=0.5M$, $\nu_U$ drops from $292.7$~Hz ($b=0$) to $291.4$~Hz ($b=M$) and $287.6$~Hz ($b=2M$). The same ordering holds in the WD and ER2--ER4 models. Because $R_{\rm ISCO}$ does not depend on $b$, the ISCO frequency is also independent of $b$ ($\nu_{\phi}^{\rm ISCO}=460.2$~Hz for $Q=0.5M$ and $l=100M$, for $b=1$, $2$ and $3M$ alike). The regularization therefore changes only the shape of the tracks above the ISCO, not their end point.

As a result, $b$ cannot be separated from $Q$ and $l$ by its effect on the ISCO frequency; any signature of $b$ must come from the shape of the $\nu_U$--$\nu_L$ relation away from the ISCO, where $b^{2}/R^{2}$ is small. This is precisely the regime probed by the data in the next subsection. The 3:2 ratio is also reached at a different radius: in the RP model it lies at $R_{3:2}=6.75M$ for Schwarzschild and moves inward to $5.57M$ for $Q=0.8M$, while for $Q=0.5M$ and $l=100M$ it moves outward from $R_{3:2}=6.08M$ ($b=0$) to $6.28M$ ($b=3M$).
\begin{table*}
\caption{\label{tab:char}Characteristic radii and frequencies for representative configurations. ``BH'' denotes the singular $b=0$ black hole, ``BB'' a black bounce with a single horizon ($R_{-}<b<R_{+}$), and ``WH'' a horizonless wormhole ($b>R_{+}$). $R_{+}$ is the outer root of $F(R)=0$; $\mathcal{E}_{\rm ISCO}$ and $\mathcal{L}_{\rm ISCO}$ are the specific energy and angular momentum at the ISCO. Frequencies are given for $M=5.4M_{\odot}$: $\nu_{\phi}^{\rm ISCO}$ is the orbital frequency at the ISCO, and $r_{3:2}$ and $\nu_{U}^{3:2}$ are the radius and upper frequency at which the RP model gives $\nu_U:\nu_L=3:2$.}
\begin{ruledtabular}
\begin{tabular}{cccccccccccc}
$Q/M$ & $b/M$ & $l/M$ & Type & $R_{+}/M$ & $R_{\rm ISCO}/M$ & $r_{\rm ISCO}/M$ & $\mathcal{E}_{\rm ISCO}$ & $\mathcal{L}_{\rm ISCO}/M$ & $\nu_{\phi}^{\rm ISCO}$ (Hz) & $r_{3:2}/M$ & $\nu_{U}^{3:2}$ (Hz) \\ \hline
0.0 & 0 & $\infty$ & BH & 2.000 & 6.000 & 6.000 & 0.9428 & 3.464 & 407.1 & 6.750 & 341.2 \\
0.0 & 0 & 100 & BH & 1.999 & 5.834 & 5.834 & 0.9478 & 3.500 & 428.8 & 6.447 & 370.5 \\
0.0 & 0 & 40 & BH & 1.995 & 5.372 & 5.372 & 0.9718 & 3.653 & 503.4 & 5.758 & 458.2 \\
0.5 & 0 & $\infty$ & BH & 1.866 & 5.607 & 5.607 & 0.9392 & 3.338 & 440.6 & 6.326 & 368.6 \\
0.8 & 0 & $\infty$ & BH & 1.600 & 4.891 & 4.891 & 0.9313 & 3.108 & 515.8 & 5.565 & 428.8 \\
0.8 & 0 & 100 & BH & 1.599 & 4.808 & 4.808 & 0.9348 & 3.128 & 531.8 & 5.402 & 451.5 \\
0.5 & 1 & 100 & BB & 1.865 & 5.475 & 5.383 & 0.9436 & 3.367 & 460.2 & 6.015 & 393.8 \\
0.5 & 2 & 100 & WH & 1.865 & 5.475 & 5.097 & 0.9436 & 3.367 & 460.2 & 5.823 & 388.3 \\
0.5 & 3 & 100 & WH & 1.865 & 5.475 & 4.580 & 0.9436 & 3.367 & 460.2 & 5.513 & 377.7 \\
\end{tabular}
\end{ruledtabular}
\end{table*}
\subsection{Statistical constraints from HF-QPOs}\label{sec:qpo}
To test the charged Simpson--Visser--AdS spacetime against observed twin-peak HF-QPOs in black-hole X-ray binaries, we use two microquasars, GRO~J1655$-$40 and XTE~J1550$-$564. We adopt the RP model, which has been widely applied to these sources and, in the Kerr metric, gives mass and spin estimates consistent with independent measurements~\cite{Motta2014MNRAS.437.2554M,Motta2014MNRAS.439L..65M}.

Throughout this subsection we work in the equatorial plane, use geometrized units with $G=c=1$, and express all lengths in units of the mass parameter $M$, so that $r$, $b$, $Q$, and $l$ are dimensionless.
\subsubsection{Data and statistical procedure}
We use the two black-hole binaries for which a twin-peak HF-QPO detection is accompanied by a dynamical mass measurement: GRO~J1655$-$40, with $\nu_{U}=450\pm3$~Hz, $\nu_{L}=300\pm5$~Hz~\cite{Strohmayer_2001ApJ...552L..49S} and $M=5.4\pm0.3\,M_{\odot}$~\cite{Beer2002MNRAS.331..351B}; and XTE~J1550$-$564, with $\nu_{U}=276\pm3$~Hz, $\nu_{L}=184\pm5$~Hz~\cite{Remillard2002ApJ...580.1030R} and $M=9.1\pm0.6\,M_{\odot}$~\cite{Orosz2011ApJ...730...75O}. For each source we define~\cite{2015EPJC...75..162B}
\begin{equation}
\chi^{2} = \sum_{i \in \{U, L\}} \frac{(\nu_{i}^{\rm mod}-\nu_{i}^{\rm obs})^{2}%
}{\sigma_{i}^2}  +\left(\frac{M-M^{\rm obs}}{\sigma_{M}}\right)^{2}.
\label{eq:chi2}
\end{equation}
The last term includes the dynamical mass as a Gaussian prior rather than as a fixed input, so each source contributes three data points. The emission radius $r_{i}$ and the mass $M_{i}$ of each source are free parameters; a fit at fixed $(Q,b,l)$ therefore has $\nu_{\rm dof}=3-2=1$ per source.

At fixed $(Q,b,l)$ the problem is two-dimensional, and we minimize $\chi^2$ by nested one-dimensional searches: for each trial radius the mass is optimized with Brent's method, and the radius is then located on a dense grid and refined in the same way. This procedure finds the global minimum of the smooth two-parameter problem without depending on a random seed. Configurations are admitted only if the orbit lies outside the ISCO ($\Omega_{r}^{2}>0$), outside the outer horizon when one exists ($R>R_{+}$), and gives $\nu_{U}>\nu_{L}>0$. We use $0\le Q\le M$ and $0\le b\le3M$ unless stated otherwise. Uncertainties are one-dimensional profile-likelihood intervals, $\Delta\chi^{2}=\chi^{2}-\chi^{2}_{\min}=1$, obtained by re-optimizing all other parameters at each value of the parameter of interest. For $l=100M$, configurations with $b<R_{-}$ are regular black holes with two horizons, those with $R_{-}<b<R_{+}$ are black bounces with a single horizon, and those with $b>R_{+}$ are traversable wormholes. Here $R_{\pm}$ are the roots of $R^{4}+l^{2}R^{2}-2l^{2}R+l^{2}Q^{2}=0$, and a horizon exists for $Q\le Q_{\rm ext}\simeq0.99995$.
\begin{table}
\caption{\label{tab:qpo}Best-fit configurations for the twin-peak HF-QPOs of GRO~J1655$-$40 and XTE~J1550$-$564 in the charged Simpson--Visser--AdS spacetime at $l=100M$. The rows with $b=M$ are black bounces with a single horizon, and those with $b=2M$ are horizonless wormholes. The emission radius $r/M$ and the mass are free parameters, with uncertainties from $\Delta\chi^{2}=1$ profile intervals. With three data points and two free parameters, $\nu_{\rm dof}=1$.}
\begin{ruledtabular}
\begin{tabular}{ccccccc}
$Q/M$ & $b/M$ & $r/M$ & $M/M_\odot$ & $\nu_U$ (Hz) & $\nu_L$ (Hz) & $\chi^2_{\rm min}$ \\ \hline
\multicolumn{7}{c}{GRO~J1655$-$40} \\ \hline
0.0 & 0.0 & $6.419\pm0.045$ & $4.488\pm0.063$ & 448.6 & 302.2 & 9.675 \\
0.0 & 1.0 & $6.355\pm0.047$ & $4.473\pm0.064$ & 448.5 & 302.3 & 9.998 \\
0.0 & 2.0 & $6.165\pm0.052$ & $4.426\pm0.066$ & 448.4 & 302.5 & 11.073 \\
0.2 & 0.0 & $6.364\pm0.045$ & $4.530\pm0.064$ & 448.6 & 302.1 & 8.804 \\
0.2 & 1.0 & $6.300\pm0.047$ & $4.515\pm0.065$ & 448.6 & 302.2 & 9.123 \\
0.2 & 2.0 & $6.108\pm0.052$ & $4.466\pm0.067$ & 448.5 & 302.5 & 10.189 \\
0.4 & 0.0 & $6.193\pm0.045$ & $4.667\pm0.067$ & 448.8 & 301.9 & 6.275 \\
0.4 & 1.0 & $6.127\pm0.047$ & $4.650\pm0.067$ & 448.7 & 302.0 & 6.575 \\
0.4 & 2.0 & $5.932\pm0.052$ & $4.596\pm0.069$ & 448.6 & 302.2 & 7.594 \\
0.6 & 0.0 & $5.887\pm0.045$ & $4.933\pm0.072$ & 449.2 & 301.3 & 2.573 \\
0.6 & 1.0 & $5.819\pm0.047$ & $4.912\pm0.073$ & 449.1 & 301.4 & 2.811 \\
0.6 & 2.0 & $5.617\pm0.053$ & $4.845\pm0.075$ & 449.0 & 301.7 & 3.655 \\
0.8 & 0.0 & $5.403\pm0.045$ & $5.416\pm0.082$ & 450.0 & 299.9 & 0.003 \\
0.8 & 1.0 & $5.331\pm0.047$ & $5.387\pm0.083$ & 450.0 & 300.0 & 0.002 \\
0.8 & 2.0 & $5.119\pm0.054$ & $5.290\pm0.086$ & 449.8 & 300.4 & 0.146 \\ \hline
\multicolumn{7}{c}{XTE~J1550$-$564} \\ \hline
0.0 & 0.0 & $6.391\pm0.071$ & $7.388\pm0.165$ & 274.2 & 186.7 & 8.805 \\
0.0 & 1.0 & $6.326\pm0.073$ & $7.367\pm0.167$ & 274.1 & 186.8 & 9.044 \\
0.0 & 2.0 & $6.130\pm0.081$ & $7.296\pm0.171$ & 274.0 & 187.1 & 9.838 \\
0.2 & 0.0 & $6.337\pm0.071$ & $7.457\pm0.167$ & 274.2 & 186.7 & 8.127 \\
0.2 & 1.0 & $6.271\pm0.073$ & $7.435\pm0.169$ & 274.2 & 186.7 & 8.366 \\
0.2 & 2.0 & $6.074\pm0.081$ & $7.362\pm0.174$ & 274.1 & 187.0 & 9.157 \\
0.4 & 0.0 & $6.168\pm0.070$ & $7.678\pm0.174$ & 274.4 & 186.4 & 6.131 \\
0.4 & 1.0 & $6.101\pm0.073$ & $7.653\pm0.176$ & 274.4 & 186.5 & 6.362 \\
0.4 & 2.0 & $5.900\pm0.081$ & $7.572\pm0.181$ & 274.3 & 186.8 & 7.137 \\
0.6 & 0.0 & $5.866\pm0.070$ & $8.104\pm0.187$ & 274.8 & 185.9 & 3.055 \\
0.6 & 1.0 & $5.797\pm0.073$ & $8.073\pm0.189$ & 274.8 & 186.0 & 3.256 \\
0.6 & 2.0 & $5.589\pm0.082$ & $7.972\pm0.195$ & 274.6 & 186.3 & 3.953 \\
0.8 & 0.0 & $5.392\pm0.071$ & $8.867\pm0.213$ & 275.7 & 184.5 & 0.173 \\
0.8 & 1.0 & $5.319\pm0.074$ & $8.822\pm0.215$ & 275.6 & 184.6 & 0.246 \\
0.8 & 2.0 & $5.101\pm0.085$ & $8.677\pm0.223$ & 275.4 & 185.0 & 0.577 \\
\end{tabular}
\end{ruledtabular}
\end{table}
\begin{table}
\caption{\label{tab:2}Profile $\chi^{2}$ in the charge parameter at $l=100M$. The single-source columns and the joint $\chi^{2}$ are computed at $b=0$, with $r$ and $M$ re-optimized at each $Q$. The last two columns give $\Delta\chi^{2}$ of the joint fit for $b=0$ and with $b$ re-optimized in $0\le b\le3M$, each measured from its own minimum ($0.130$ and $0.122$). $\Delta\chi^{2}=1$, $2.71$ and $6.63$ correspond to the $68\%$, $90\%$ and $99\%$ intervals for one parameter.}
\begin{ruledtabular}
\begin{tabular}{cccccc}
 & GRO & XTE & \multicolumn{3}{c}{Joint} \\
$Q/M$ & $\chi^{2}$ & $\chi^{2}$ & $\chi^{2}$ & $\Delta\chi^{2}_{b=0}$ & $\Delta\chi^{2}_{b\,{\rm free}}$ \\ \hline
0.00 & 9.675 & 8.805 & 18.480 & 18.350 & 18.358 \\
0.20 & 8.804 & 8.127 & 16.931 & 16.801 & 16.809 \\
0.40 & 6.275 & 6.131 & 12.406 & 12.276 & 12.284 \\
0.50 & 4.504 & 4.693 & 9.197 & 9.067 & 9.075 \\
0.60 & 2.573 & 3.055 & 5.629 & 5.499 & 5.507 \\
0.70 & 0.818 & 1.404 & 2.222 & 2.092 & 2.100 \\
0.75 & 0.215 & 0.692 & 0.907 & 0.777 & 0.785 \\
0.80 & 0.003 & 0.173 & 0.176 & 0.046 & 0.054 \\
0.85 & 0.459 & 0.006 & 0.465 & 0.336 & 0.005 \\
0.90 & 2.048 & 0.454 & 2.502 & 2.373 & 0.001 \\
0.95 & 5.595 & 1.977 & 7.572 & 7.442 & 0.250 \\
0.99 & 10.860 & 4.523 & 15.383 & 15.253 & 1.586 \\
\end{tabular}
\end{ruledtabular}
\end{table}
\begin{table}\caption{\label{tab:l}
Profile $\chi^{2}$ in the AdS radius $l$ at $Q=b=0$, with $r$ and $M$ re-optimized for each source. $\Delta\chi^{2}$ is measured from the joint minimum, $\chi^{2}_{\min}=0.138$ at $l=40.6M$. The last row is the asymptotically flat Schwarzschild limit.}
\begin{ruledtabular}
\begin{tabular}{ccccc}
 & GRO & XTE & \multicolumn{2}{c}{Joint} \\
$l/M$ & $\chi^2$ & $\chi^2$ & $\chi^2$ & $\Delta\chi^2$ \\ \hline
20 & 36.968 & 19.380 & 56.349 & 56.211 \\
30 & 5.283 & 1.877 & 7.160 & 7.022 \\
40 & 0.103 & 0.047 & 0.151 & 0.013 \\
40.6 & 0.054 & 0.084 & 0.138 & 0.000 \\
50 & 0.764 & 1.388 & 2.152 & 2.014 \\
70 & 4.859 & 5.035 & 9.893 & 9.756 \\
100 & 9.675 & 8.805 & 18.480 & 18.342 \\
200 & 15.378 & 13.039 & 28.417 & 28.279 \\
1000 & 17.851 & 14.825 & 32.676 & 32.538 \\
$\infty$ & 17.964 & 14.905 & 32.869 & 32.731 \\
\end{tabular}
\end{ruledtabular}
\end{table}
\subsubsection{Results}
We begin with the asymptotically flat Schwarzschild limit, $Q=b=0$ and $l\to\infty$. The best fit gives $r=6.697M$ and $M=4.161\,M_{\odot}$ for GRO~J1655$-$40, and $r=6.651M$ and $M=6.885\,M_{\odot}$ for XTE~J1550$-$564, with $\chi^{2}=17.96$ and $14.91$, respectively, for one degree of freedom ($p=2.3\times10^{-5}$ and $1.1\times10^{-4}$). Combined, $\chi^{2}=32.87$ for two degrees of freedom ($p=7.3\times10^{-8}$). These small $p$-values reflect a real tension between the observed 3:2 frequency ratio and the dynamical masses: in the Schwarzschild metric the ratio is reached at $R=6.75M$, and matching the observed $\nu_U$ there requires masses well below the measured ones. This is a known feature of the RP model; in the Kerr metric it is resolved by a moderate spin~\cite{Motta2014MNRAS.437.2554M,Motta2014MNRAS.439L..65M}.

Table~\ref{tab:qpo} lists the best-fit configurations on a grid in $(Q/M,b/M)$ at fixed $l=100M$. Two trends are clear. A nonzero $Q$ steadily improves the fit for both sources, and by $Q=0.8M$ both pairs and both masses are reproduced. At fixed $Q$, a larger $b$ makes the fit slightly worse, because at a given ratio $\nu_L/\nu_U$ the regularization moves the emission radius outward in $R$ and so lowers the inferred mass. A joint fit in which the two sources share a single value of $Q$ at $b=0$ has six data points and five free parameters ($r_{1,2}$, $M_{1,2}$, $Q$), so $\nu_{\rm dof}=1$, and gives
\begin{equation}
Q/M=0.814^{+0.059}_{-0.074}, \qquad  \chi^{2}_{\rm min}=0.130\, .
\label{eq:jointQ}
\end{equation}
At this point the best-fit emission radii are $r=5.358M$ and $5.349M$, and the masses are $5.46\,M_{\odot}$ and $8.94\,M_{\odot}$ for GRO~J1655$-$40 and XTE~J1550$-$564, respectively, both consistent with the dynamical values. Both emission radii lie outside the ISCO ($r_{\rm ISCO}=4.762M$) and the outer horizon ($R_{+}=1.580M$), and the configuration is sub-extremal ($Q_{\rm ext}\simeq0.99995M$ at $l=100M$).

\begin{figure}
\centering
\includegraphics[width=0.95\columnwidth]{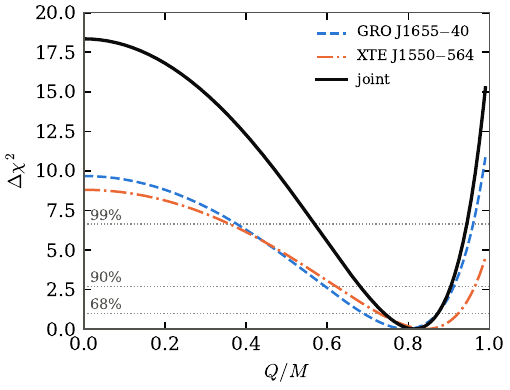}
\caption{$\Delta\chi^{2}$ as a function of $Q/M$ at $b=0$ and $l=100M$, for the joint fit of GRO~J1655$-$40 and XTE~J1550$-$564 (solid) and for each source separately (dashed and dot-dashed, measured from each source's own minimum). The dotted lines mark $\Delta\chi^{2}=1$, $2.71$ and $6.63$ (68\%, 90\% and 99\% for one parameter). At this fixed $l$, the joint minimum lies at $Q/M=0.814^{+0.059}_{-0.074}$, and $Q=0$ is disfavored by $\Delta\chi^{2}=18.35$.}\label{fig:dchi2}
\end{figure}
Figure~\ref{fig:dchi2} shows the joint profile $\Delta\chi^{2}(Q)$; at each $Q$ the emission radius and mass of each source are re-optimized independently. The two sources individually prefer $Q/M\simeq0.80$ (GRO~J1655$-$40) and $0.85$ (XTE~J1550$-$564) and are mutually consistent; the joint minimum is well defined. The $90\%$ and $99\%$ intervals ($\Delta\chi^{2}=2.71$ and $6.63$) are $0.680\le Q/M\le0.905$ and $0.569\le Q/M\le0.944$. The curve rises toward $Q=0$, where $\chi^{2}=18.48$, i.e., $\Delta\chi^{2}=18.35$ above the minimum. Taken at face value this corresponds to $4.3\sigma$ for one parameter, or $4.4\sigma$ if one accounts for $Q=0$ lying on the boundary of the parameter $Q^{2}\ge0$. Near the minimum the profile is shallower on the low-$Q$ side, because the frequencies depend on $Q^{2}$, and a given change in $Q$ has less effect at small $Q$ than near extremality. Toward $Q\to Q_{\rm ext}$ the profile rises steeply again.

\begin{figure}
\centering
\includegraphics[width=0.95\columnwidth]{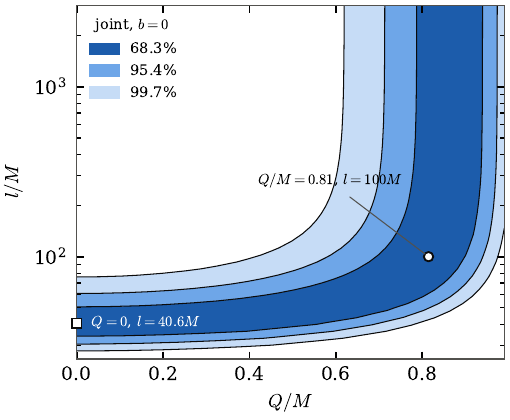}
\caption{Joint confidence regions in the $(Q/M,\,l/M)$ plane at $b=0$, with the emission radius and mass of each source re-optimized at every point. The shaded regions correspond to $\Delta\chi^{2}=2.30$, $6.18$ and $11.83$ (68.3\%, 95.4\% and 99.7\% for two parameters). The circle marks the charged fit of Eq.~\eqref{eq:jointQ} at $l=100M$, and the square marks the uncharged fit of Eq.~\eqref{eq:jointL}.}\label{fig:Ql}
\end{figure}

This preference for nonzero $Q$ holds only at fixed $l$. Table~\ref{tab:l} profiles $\chi^{2}$ over $l$ at $Q=b=0$. The fit has a pronounced minimum,
\begin{equation}
l/M=40.6^{+6.2}_{-4.8}, \qquad \chi^{2}_{\rm min}=0.138 ,
\label{eq:jointL}
\end{equation}
with emission radii $r=5.774M$ and $5.770M$ and masses $5.47\,M_{\odot}$ and $8.93\,M_{\odot}$. This uncharged fit is as good as the charged fit of Eq.~\eqref{eq:jointQ}. A smaller $l$ ($l\to20M$) confines the orbits too strongly, while a larger $l$ ($l\to1000M$) relaxes back toward the disfavored Schwarzschild value. Figure~\ref{fig:Ql} maps the joint $\Delta\chi^{2}$ over the full $(Q,l)$ plane. The allowed region is a single curved valley that connects the uncharged solution at $l\simeq40M$ with charged solutions at larger $l$. In the asymptotically flat limit $l\to\infty$ it ends at $Q/M=0.875^{+0.045}_{-0.055}$ ($\chi^{2}_{\min}=0.126$). The minimum $\chi^{2}$ is essentially the same everywhere along the valley, so the data fix one combination of $Q$ and $l$, not both.

The regularization parameter is even less constrained. Repeating the joint fit at $l=100M$ with $Q$ free for fixed values of $b$, we find that $\chi^{2}_{\min}$ changes only from $0.130$ ($b=0$) to $0.119$ ($b=3.5M$), while the preferred charge rises from $0.814M$ to $0.983M$. Larger values of $b$ would need $Q>M$, outside the range we allow, and the fit then degrades: $\Delta\chi^{2}=1$ is reached at $b\simeq3.9M$ and $\Delta\chi^{2}=2.71$ at $b\simeq4.2M$. This upper limit on $b$ is therefore set by the charge prior, not by the data themselves. The last column of Table~\ref{tab:2} shows the same degeneracy from the other side: once $b$ is free, charges up to $Q\simeq0.95M$ stay within $\Delta\chi^{2}<1$ of the best fit.

Finally, we compare the deformed spacetimes with the Kerr metric. In the RP model a Kerr black hole reproduces each pair exactly at the dynamical mass, with spin $a\simeq0.34$ (GRO~J1655$-$40; $0.28$--$0.40$ for $M$ within $\pm1\sigma$) and $a\simeq0.38$ (XTE~J1550$-$564; $0.30$--$0.44$). These values are close to the RP estimates obtained from the full QPO triplets, $a=0.290\pm0.003$ and $a=0.34\pm0.01$~\cite{Motta2014MNRAS.437.2554M,Motta2014MNRAS.439L..65M}. In terms of the Akaike information criterion, ${\rm AIC}=\chi^{2}_{\min}+2k$~\cite{Akaike1974ITAC...19..716A}, with $k$ free parameters, the Schwarzschild fit ($k=4$) gives ${\rm AIC}=40.9$. Either one-parameter deformation ($Q$ at fixed $l$, or $l$ at $Q=0$; $k=5$) gives ${\rm AIC}=10.1$, and the Kerr description with one spin per source ($k=6$) gives $12.0$. The data thus clearly require a departure from Schwarzschild, but they do not prefer any particular deformation.
\subsubsection{Degeneracies and interpretation}
Several degeneracies limit what these fits can tell us. The first is between $Q$ and $l$. At fixed $R$, both a larger $Q$ and a smaller $l$ raise $\Omega_{r}/\Omega_{\phi}$ (Sec.~\ref{subsec: Harmonic_Oscs}), so the 3:2 ratio is reached closer to the black hole, where $\Omega_{\phi}$ is higher. This in turn raises the mass inferred from $\nu_{U}$. Two measured frequencies and a mass per source can fix this shift, but not how it is shared between $Q$ and $l$. Accordingly, the value in Eq.~\eqref{eq:jointQ} holds only for the slice $l=100M$, and Eq.~\eqref{eq:jointL} gives an equally good description without any charge.

The second degeneracy involves $b$. By Eq.~\eqref{eq:Omr_R}, $b$ enters only through the factor $1-b^{2}/R^{2}$ in $\Omega_{r}^{2}$, which acts in the opposite direction to $Q$. At the relevant emission radii, $R\simeq5.3$--$6.4M$, this factor differs from unity by about $0.03$ for $b=M$ and $0.2$--$0.3$ for $b=3M$, and a slightly larger charge compensates it fully.

The third, and probably the most important, involves the spin. The metric~\eqref{metric} is static, whereas both sources have independently estimated spins. Prograde frame dragging raises $\nu_{r}/\nu_{\phi}$ at fixed radius, exactly as a larger $Q$ or a smaller $l$ does, and the Kerr fits above show that a modest spin suffices. The deformation found here therefore most likely absorbs the neglected rotation in addition to (or instead of) any genuine charge or AdS curvature. It should be read as an effective measure of the departure from Schwarzschild rather than as an estimate of each parameter separately.

The physical scales involved support this reading. The value $l\simeq40M$ corresponds to $l\approx330$~km for GRO~J1655$-$40 and $\approx540$~km for XTE~J1550$-$564, i.e., $|\Lambda|=3/l^{2}\sim10^{-11}\,{\rm m^{-2}}$, some $41$ orders of magnitude above the observed cosmological constant (and of the opposite sign). Likewise, an astrophysical black hole is not expected to retain an electric charge $Q/M\simeq0.8$ in a plasma environment. Moreover, as noted below Eq.~\eqref{nu_i}, the conversion of the Killing-time frequencies to observed ones is uncertain at the level of $R^{2}/l^{2}$, which for $l\simeq40M$ amounts to about $2\%$ at $R\simeq6M$. This is comparable to the measurement errors, so the AdS term can only be viewed as an effective, local deformation of the near-zone geometry.

Breaking these degeneracies requires a rotating generalization of the metric, or additional timing information such as a genuine low-frequency (nodal-precession) QPO, which the present static geometry cannot generate. In summary, physically admissible configurations of the charged Simpson--Visser--AdS spacetime reproduce both observed HF-QPO pairs and the dynamical masses, and the data require a departure from the Schwarzschild limit; the individual parameters $b$, $Q$ and $l$, however, remain degenerate. The entries of Tables~\ref{tab:qpo}--\ref{tab:l} should be read as allowed configurations subject to these degeneracies, not as independent determinations of the spacetime parameters.
\section{Concluding remarks}\label{sec:conclusion}
In this article, we considered the charged Simpson--Visser--AdS spacetime and examined its horizon structure, circular motion, fundamental frequencies, and twin-peak QPOs. Our main results can be summarized as follows.

The roots of $f(r)=0$ divide the parameter space into regular black holes with two horizons ($b<R_{-}$), black bounces with a single horizon ($R_{-}<b<R_{+}$), and horizonless traversable wormholes ($b>R_{+}$, or $Q>Q_{\rm ext}$ with $b>0$), with the one-way wormhole $b=R_{+}$ as the borderline case. Since $Q$ and $b$ both reduce the outer horizon radius, together they shrink the region in which black holes exist. Increasing $l$ weakens the AdS confinement and lets $r_{+}$ approach its asymptotically flat value.

Writing the metric in terms of the areal radius $R=\sqrt{r^{2}+b^{2}}$ turned out to be the key to the circular-orbit problem. At fixed $R$, the effective potential, the specific energy and angular momentum of circular orbits, the Keplerian frequency, and the ISCO are exactly those of the RN-AdS black hole, and we gave the ISCO condition in closed form [Eq.~\eqref{eq:isco_poly}]. The charge moves the ISCO inward and lowers $\mathcal{E}_{\rm ISCO}$ and $\mathcal{L}_{\rm ISCO}$; a smaller AdS radius also moves the ISCO inward but raises $\mathcal{E}_{\rm ISCO}$ and $\mathcal{L}_{\rm ISCO}$, and for $l\gtrsim500M$ the results are close to those of the asymptotically flat charged Simpson--Visser black hole. The regularization parameter reduces the coordinate radius $r_{\rm ISCO}=\sqrt{R_{\rm ISCO}^{2}-b^{2}}$, but leaves $R_{\rm ISCO}$, $\mathcal{E}_{\rm ISCO}$, $\mathcal{L}_{\rm ISCO}$, and the orbital frequency at the ISCO unchanged.

In the epicyclic spectrum, $b$ appears only through the factor $1-b^{2}/R^{2}$ in $\Omega_{r}^{2}$ [Eq.~\eqref{eq:Omr_R}]. At fixed radius, the charge lowers the Keplerian frequency and raises the radial one, a smaller $l$ raises both, and $b$ lowers the radial frequency. In AdS, $\Omega_{\phi}\to1/l$ and $\Omega_{r}\to2/l$ at large radii, so the RP lower frequency becomes negative beyond a radius that depends on $l$. In the $\nu_{U}$--$\nu_{L}$ plane, $Q$ and a smaller $l$ shift the tracks of all five QPO models to higher $\nu_{U}$ at a given $\nu_{L}$ and raise the ISCO frequency. The regularization shifts the tracks slightly in the opposite direction but does not move their ISCO end point.

Within the RP framework, the Schwarzschild limit cannot reproduce the HF-QPO pairs of GRO~J1655$-$40 and XTE~J1550$-$564 together with their dynamical masses ($\chi^{2}=32.9$ for two degrees of freedom). A single additional deformation removes this tension. At fixed $l=100M$, the joint fit gives $Q/M=0.814^{+0.059}_{-0.074}$, but an uncharged spacetime with $l=40.6^{+6.2}_{-4.8}M$ fits equally well, and both solutions lie on one degeneracy valley in the $(Q,l)$ plane. The regularization parameter is unconstrained up to $b\simeq3.9M$, a limit set by our prior $Q\le M$. A Kerr black hole with $a\simeq0.3$--$0.4$ describes the same data, and the required values of $l$ and $Q$ are far from anything expected for an astrophysical black hole. We therefore read the preferred deformation as an effective measure of the departure from Schwarzschild, which most likely absorbs the spin neglected here, rather than as a detection of charge, AdS curvature or regularization. Disentangling these effects will require a rotating generalization of the spacetime, a larger sample of sources, or additional timing information, such as a genuine low-frequency nodal-precession QPO beyond the two HF-QPO peaks considered in this work.
\section*{Acknowledgment}
J.R. thanks Grant No. FA-F-2021-510 of the Uzbekistan Agency for Innovative Development and the Silesian University of Opava for hospitality. The work of Weiwei Wang was supported by the NSF of Fujian Province of China (Grant Nos. 2024J011011 and 2022J01105).
\subsection*{Declaration of competing interest}
The authors declare that they have no known competing financial interests or personal relationships that could have appeared to influence the work reported in this paper.
\subsection*{Data availability}
The QPO frequencies and masses used in this work are taken from the cited literature. The code used for the fits is available from the corresponding author upon reasonable request.
\def\prc{Phys. Rev. C}
\def\nar{New Astronomy Reviews}
\def\pre{Phys. Rev. E}
\def\prd{Phys. Rev. D}
\def\prl{Physical Review Letters}
\def\jcap{Journal of Cosmology and Astroparticle Physics}
\def\apss{Astrophysics and Space Science}
\def\mnras{Monthly Notices of the Royal Astronomical Society}
\def\apj{The Astrophysical Journal}
\def\aap{Astronomy and Astrophysics}
\def\actaa{Acta Astronomica}
\def\pasj{Publications of the Astronomical Society of Japan}
\def\apjl{Astrophysical Journal Letters}
\def\pasa{Publications Astronomical Society of Australia}
\def\nat{Nature}
\def\physrep{Physics Reports}
\def\araa{Annual Review of Astronomy and Astrophysics}
\def\apjs{The Astrophysical Journal Supplement}
\def\aapr{The Astronomy and Astrophysics Review}
\def\procspie{Proceedings of the SPIE}

\def\prd{Phys. Rev. D}
\def\prl{Physical Review Letters}
\def\jcap{Journal of Cosmology and Astroparticle Physics}
\def\mnras{Monthly Notices of the Royal Astronomical Society}
\def\apj{The Astrophysical Journal}
\def\aap{Astronomy and Astrophysics}
\def\apjl{Astrophysical Journal Letters}
\def\araa{Annual Review of Astronomy and Astrophysics}
\def\newar{New Astronomy Reviews}
\bibliographystyle{apsrev4-2}
\bibliography{References}

@ARTICLE{Khan2026EPJC...86..597K,
       author = {{Khan}, Saeed Ullah and {Rayimbaev}, Javlon and {Hayat}, Haidar and {Chen}, Zhi-Min and {Abdullaev}, Mardon and {Murodov}, Sardor and {Wang}, Weiwei},
        title = "{Quasi-periodic oscillations in rotating Simpson─Visser black holes in STVG}",
      journal = {European Physical Journal C},
         year = 2026,
        month = jun,
       volume = {86},
       number = {6},
          eid = {597},
        pages = {597},
          doi = {10.1140/epjc/s10052-026-15830-w},
       adsurl = {https://ui.adsabs.harvard.edu/abs/2026EPJC...86..597K}
}

@ARTICLE{Khan2026ChJPh.102..711K,
       author = {{Khan}, Saeed Ullah and {Zahid}, Muhammad and {Chen}, Zhi-Min and {Turaev}, Yunus and {Akhmedov}, Munisbek and {Usanov}, Sulton and {Rayimbaev}, Javlon and {Wang}, Weiwei},
        title = "{Rotating black-bounce black holes in modified gravity: Photon orbits and shadow}",
      journal = {Chinese Journal of Physics},
         year = 2026,
        month = aug,
       volume = {102},
        pages = {711-727},
          doi = {10.1016/j.cjph.2026.03.023},
       adsurl = {https://ui.adsabs.harvard.edu/abs/2026ChJPh.102..711K}
}

@ARTICLE{Simpson2019JCAP...02..042S,
       author = {{Simpson}, Alex and {Visser}, Matt},
        title = "{Black-bounce to traversable wormhole}",
      journal = {\jcap},
         year = 2019,
        month = feb,
       volume = {2019},
       number = {2},
          eid = {042},
        pages = {042},
          doi = {10.1088/1475-7516/2019/02/042},
archivePrefix = {arXiv},
       eprint = {1812.07114},
 primaryClass = {gr-qc},
       adsurl = {https://ui.adsabs.harvard.edu/abs/2019JCAP...02..042S}
}

@ARTICLE{2021JCAP...04..082M,
       author = {{Mazza}, Jacopo and {Franzin}, Edgardo and {Liberati}, Stefano},
        title = "{A novel family of rotating black hole mimickers}",
      journal = {\jcap},
         year = 2021,
        month = apr,
       volume = {2021},
       number = {4},
          eid = {082},
        pages = {082},
          doi = {10.1088/1475-7516/2021/04/082},
archivePrefix = {arXiv},
       eprint = {2102.01105},
 primaryClass = {gr-qc},
       adsurl = {https://ui.adsabs.harvard.edu/abs/2021JCAP...04..082M}
}

@ARTICLE{2021PhRvL.126j1102B,
       author = {{Bl{\'a}zquez-Salcedo}, Jose Luis and {Knoll}, Christian and {Radu}, Eugen},
        title = "{Traversable Wormholes in Einstein-Dirac-Maxwell Theory}",
      journal = {\prl},
         year = 2021,
        month = mar,
       volume = {126},
       number = {10},
          eid = {101102},
        pages = {101102},
          doi = {10.1103/PhysRevLett.126.101102},
archivePrefix = {arXiv},
       eprint = {2010.07317},
 primaryClass = {gr-qc},
       adsurl = {https://ui.adsabs.harvard.edu/abs/2021PhRvL.126j1102B}
}

@ARTICLE{2022EPJC...82..533B,
       author = {{Bl{\'a}zquez-Salcedo}, Jose Luis and {Knoll}, Christian and {Radu}, E.},
        title = "{Einstein-Dirac-Maxwell wormholes: ansatz, construction and properties of symmetric solutions}",
      journal = {European Physical Journal C},
         year = 2022,
        month = jun,
       volume = {82},
       number = {6},
          eid = {533},
        pages = {533},
          doi = {10.1140/epjc/s10052-022-10488-6},
archivePrefix = {arXiv},
       eprint = {2108.12187},
 primaryClass = {gr-qc},
       adsurl = {https://ui.adsabs.harvard.edu/abs/2022EPJC...82..533B}
}

@ARTICLE{2022PhRvL.128i1104K,
       author = {{Konoplya}, R.~A. and {Zhidenko}, A.},
        title = "{Traversable Wormholes in General Relativity}",
      journal = {\prl},
         year = 2022,
        month = mar,
       volume = {128},
       number = {9},
          eid = {091104},
        pages = {091104},
          doi = {10.1103/PhysRevLett.128.091104},
archivePrefix = {arXiv},
       eprint = {2106.05034},
 primaryClass = {gr-qc},
       adsurl = {https://ui.adsabs.harvard.edu/abs/2022PhRvL.128i1104K}
}

@ARTICLE{2023EPJC...83..854V,
       author = {{Vrba}, Jaroslav and {Rayimbaev}, Javlon and {Stuchlik}, Zdenek and {Ahmedov}, Bobomurat},
        title = "{Charged particles motion and quasiperiodic oscillation in Simpson─Visser spacetime in the presence of external magnetic fields}",
      journal = {European Physical Journal C},
         year = 2023,
        month = sep,
       volume = {83},
       number = {9},
          eid = {854},
        pages = {854},
          doi = {10.1140/epjc/s10052-023-12023-7},
       adsurl = {https://ui.adsabs.harvard.edu/abs/2023EPJC...83..854V}
}

@ARTICLE{Ahmed2026EPJC...86..658A,
       author = {{Ahmed}, Faizuddin and {Al-Badawi}, Ahmad and {Fathi}, Mohsen},
        title = "{Charged Simpson─Visser AdS black holes: geodesic structure and thermodynamic properties}",
      journal = {European Physical Journal C},
         year = 2026,
        month = jun,
       volume = {86},
       number = {6},
          eid = {658},
        pages = {658},
          doi = {10.1140/epjc/s10052-026-15935-2},
archivePrefix = {arXiv},
       eprint = {2601.10469},
 primaryClass = {gr-qc},
       adsurl = {https://ui.adsabs.harvard.edu/abs/2026EPJC...86..658A}
}

@ARTICLE{2025PhRvD.112l4018D,
       author = {{Dasgupta}, Anirban and {Banerjee}, Indrani},
        title = "{Constraining the rotating Simpson-Visser spacetime from the observed quasiperiodic oscillations in black holes}",
      journal = {\prd},
         year = 2025,
        month = dec,
       volume = {112},
       number = {12},
          eid = {124018},
        pages = {124018},
          doi = {10.1103/2x2n-t383},
archivePrefix = {arXiv},
       eprint = {2509.15761},
 primaryClass = {gr-qc},
       adsurl = {https://ui.adsabs.harvard.edu/abs/2025PhRvD.112l4018D}
}

@ARTICLE{Ingram2010MNRAS,
       author = {{Ingram}, Adam and {Done}, Chris},
        title = "{A physical interpretation of the variability power spectral components in accreting neutron stars}",
      journal = {Mon.Not.R.Astron.Soc},
         year = 2010,
        month = jul,
       volume = {405},
       number = {4},
        pages = {2447-2452},
          doi = {10.1111/j.1365-2966.2010.16614.x},
archivePrefix = {arXiv},
       eprint = {0907.5485},
 primaryClass = {astro-ph.SR},
       adsurl = {https://ui.adsabs.harvard.edu/abs/2010MNRAS.405.2447I}
}

@BOOK{Chandrasekhar1983mtbh.book.C,
       author = {{Chandrasekhar}, S.},
        title = "{The mathematical theory of black holes}",
         year = 1983,
       adsurl = {https://ui.adsabs.harvard.edu/abs/1983mtbh.book.....C}, 
      publisher = {Oxford University Press}
}

@BOOK{Bambi2017bhlt.book..B,
       author = {{Bambi}, Cosimo},
        title = "{Black Holes: A Laboratory for Testing Strong Gravity}",
         year = 2017,
          doi = {10.1007/978-981-10-4524-0},
       adsurl = {https://ui.adsabs.harvard.edu/abs/2017bhlt.book.....B},
      publisher = {Singapore: Springer}
}

@ARTICLE{Zdunik2000AA,
       author = {{Zdunik}, J.~L. and {Haensel}, P. and {Gondek-Rosi{\'n}ska}, D. and {Gourgoulhon}, E.},
        title = "{Innermost stable circular orbits around strange stars and kHz QPOs in low-mass X-ray binaries}",
      journal = {Astron.Astrophys.},
         year = 2000,
        month = apr,
       volume = {356},
        pages = {612-618},
archivePrefix = {arXiv},
       eprint = {astro-ph/0002394},
 primaryClass = {astro-ph},
       adsurl = {https://ui.adsabs.harvard.edu/abs/2000A&A...356..612Z}
}

@ARTICLE{Wilkins2012MNRAS,
       author = {{Wilkins}, D.~R. and {Fabian}, A.~C.},
        title = "{Understanding X-ray reflection emissivity profiles in AGN: locating the X-ray source}",
      journal = {\mnras},
         year = 2012,
        month = aug,
       volume = {424},
       number = {2},
        pages = {1284-1296},
          doi = {10.1111/j.1365-2966.2012.21308.x},
archivePrefix = {arXiv},
       eprint = {1205.3179},
 primaryClass = {astro-ph.HE},
       adsurl = {https://ui.adsabs.harvard.edu/abs/2012MNRAS.424.1284W}
}

@ARTICLE{Aliev2002MNRAS.336..241A,
       author = {{Aliev}, A.~N. and {{\"O}zdemir}, N.},
        title = "{Motion of charged particles around a rotating black hole in a magnetic field}",
      journal = {\mnras},
         year = 2002,
        month = oct,
       volume = {336},
       number = {1},
        pages = {241-248},
          doi = {10.1046/j.1365-8711.2002.05727.x},
archivePrefix = {arXiv},
       eprint = {gr-qc/0208025},
 primaryClass = {gr-qc},
       adsurl = {https://ui.adsabs.harvard.edu/abs/2002MNRAS.336..241A}
}

@ARTICLE{Vrba2020PhRvD.101l4039V,
       author = {{Vrba}, Jaroslav and {Abdujabbarov}, Ahmadjon and {Kolo{\v{s}}}, Martin and {Ahmedov}, Bobomurat and {Stuchl{\'\i}k}, Zden{\v{e}}k and {Rayimbaev}, Javlon},
        title = "{Charged and magnetized particles motion in the field of generic singular black holes governed by general relativity coupled to nonlinear electrodynamics}",
      journal = {\prd},
         year = 2020,
        month = jun,
       volume = {101},
       number = {12},
          eid = {124039},
        pages = {124039},
          doi = {10.1103/PhysRevD.101.124039},
       adsurl = {https://ui.adsabs.harvard.edu/abs/2020PhRvD.101l4039V}
}

@ARTICLE{Zahid2022EPJC...82..494Z,
       author = {{Zahid}, Muhammad and {Rayimbaev}, Javlon and {Khan}, Saeed Ullah and {Ren}, Jingli and {Ahmedov}, Saidmuhammad and {Ibragimov}, Inomjon},
        title = "{Dynamics and collisions of magnetized particles around charged black holes in Einstein-Maxwell-scalar theory}",
      journal = {European Physical Journal C},
         year = 2022,
        month = may,
       volume = {82},
       number = {5},
          eid = {494},
        pages = {494},
          doi = {10.1140/epjc/s10052-022-10432-8},
       adsurl = {https://ui.adsabs.harvard.edu/abs/2022EPJC...82..494Z}
}

@ARTICLE{Amendola2008JCAP...04..013A,
       author = {{Amendola}, Luca and {Kunz}, Martin and {Sapone}, Domenico},
        title = "{Measuring the dark side (with weak lensing)}",
      journal = {\jcap},
         year = 2008,
        month = apr,
       volume = {2008},
       number = {4},
          eid = {013},
        pages = {013},
          doi = {10.1088/1475-7516/2008/04/013},
archivePrefix = {arXiv},
       eprint = {0704.2421},
 primaryClass = {astro-ph},
       adsurl = {https://ui.adsabs.harvard.edu/abs/2008JCAP...04..013A}
}

@ARTICLE{2025EPJC...85.1029N,
       author = {{Nishonov}, Isomiddin and {Murodov}, Sardor and {Ahmedov}, Bobomurat and {Khan}, Saeed Ullah and {Rayimbaev}, Javlon and {Ibragimov}, Inomjon and {Sabirov}, Sardor},
        title = "{QPOs from charged particles around charged black holes in STVG}",
      journal = {European Physical Journal C},
         year = 2025,
        month = sep,
       volume = {85},
       number = {9},
          eid = {1029},
        pages = {1029},
          doi = {10.1140/epjc/s10052-025-14751-4},
       adsurl = {https://ui.adsabs.harvard.edu/abs/2025EPJC...85.1029N}
}

@ARTICLE{2021EPJC...81.1067S,
       author = {{Shahzadi}, Misbah and {Kolo{\v{s}}}, Martin and {Stuchl{\'\i}k}, Zden{\v{e}}k and {Habib}, Yousaf},
        title = "{Epicyclic oscillations in spinning particle motion around Kerr black hole applied in models fitting the quasi-periodic oscillations observed in microquasars and AGNs}",
      journal = {European Physical Journal C},
         year = 2021,
        month = dec,
       volume = {81},
       number = {12},
          eid = {1067},
        pages = {1067},
          doi = {10.1140/epjc/s10052-021-09868-1},
archivePrefix = {arXiv},
       eprint = {2104.09640},
 primaryClass = {astro-ph.HE},
       adsurl = {https://ui.adsabs.harvard.edu/abs/2021EPJC...81.1067S}
}

@ARTICLE{Stella1999ApJ,
       author = {{Stella}, Luigi and {Vietri}, Mario and {Morsink}, Sharon M.},
        title = "{Correlations in the Quasi-periodic Oscillation Frequencies of Low-Mass X-Ray Binaries and the Relativistic Precession Model}",
      journal = {The Astrophysical Journal},
         year = 1999,
        month = oct,
       volume = {524},
       number = {1},
        pages = {L63-L66},
          doi = {10.1086/312291},
archivePrefix = {arXiv},
       eprint = {astro-ph/9907346},
 primaryClass = {astro-ph},
       adsurl = {https://ui.adsabs.harvard.edu/abs/1999ApJ...524L..63S}
}

@ARTICLE{EHT2019ApJ...875L...1E,
       author = {{Event Horizon Telescope Collaboration}},
        title = "{First M87 Event Horizon Telescope Results. I. The Shadow of the Supermassive Black Hole}",
      journal = {\apjl},
         year = 2019,
        month = apr,
       volume = {875},
       number = {1},
          eid = {L1},
        pages = {L1},
          doi = {10.3847/2041-8213/ab0ec7},
archivePrefix = {arXiv},
       eprint = {1906.11238},
 primaryClass = {astro-ph.GA},
       adsurl = {https://ui.adsabs.harvard.edu/abs/2019ApJ...875L...1E}
}

@ARTICLE{EHT2023ApJ...957L..20E,
       author = {{Event Horizon Telescope Collaboration} },
        title = "{First M87 Event Horizon Telescope Results. IX. Detection of Near-horizon Circular Polarization}",
      journal = {\apjl},
         year = 2023,
        month = nov,
       volume = {957},
       number = {2},
          eid = {L20},
        pages = {L20},
          doi = {10.3847/2041-8213/acff70},
archivePrefix = {arXiv},
       eprint = {2311.10976},
 primaryClass = {astro-ph.HE},
       adsurl = {https://ui.adsabs.harvard.edu/abs/2023ApJ...957L..20E}
}

@ARTICLE{EHT2022ApJ...930L..17E,
       author = {{Event Horizon Telescope Collaboration}},
        title = "{First Sagittarius A* Event Horizon Telescope Results. VI. Testing the Black Hole Metric}",
      journal = {\apjl},
         year = 2022,
        month = may,
       volume = {930},
       number = {2},
          eid = {L17},
        pages = {L17},
          doi = {10.3847/2041-8213/ac6756},
archivePrefix = {arXiv},
       eprint = {2311.09484},
 primaryClass = {astro-ph.HE},
       adsurl = {https://ui.adsabs.harvard.edu/abs/2022ApJ...930L..17E}
}

@ARTICLE{EHT2022ApJ...930L..12E,
       author = {{Event Horizon Telescope Collaboration}},
        title = "{First Sagittarius A* Event Horizon Telescope Results. I. The Shadow of the Supermassive Black Hole in the Center of the Milky Way}",
      journal = {\apjl},
         year = 2022,
        month = may,
       volume = {930},
       number = {2},
          eid = {L12},
        pages = {L12},
          doi = {10.3847/2041-8213/ac6674},
archivePrefix = {arXiv},
       eprint = {2311.08680},
 primaryClass = {astro-ph.HE},
       adsurl = {https://ui.adsabs.harvard.edu/abs/2022ApJ...930L..12E}
}

@article{khan2024circular,
  title={Circular motion and collisions of particles with magnetic dipole moment and electric charge in dipolar magnetosphere around Schwarzschild black holes},
  author={Khan, Saeed Ullah and Abdurkhmonov, Ozodbek and Rayimbaev, Javlon and Ahmedov, Saidmuhammad and Turaev, Yunus and Muminov, Sokhibjan},
  journal={The European Physical Journal C},
  volume={84},
  number={6},
  pages={650},
  year={2024},
  publisher={Springer}
}

@ARTICLE{2022ChJPh..78..141K,
       author = {{Khan}, Saeed Ullah and {Ren}, Jingli},
        title = "{Geodesics and optical properties of a rotating black hole in Randall-Sundrum brane with a cosmological constant}",
      journal = {Chinese Journal of Physics},
         year = 2022,
        month = aug,
       volume = {78},
        pages = {141-154},
          doi = {10.1016/j.cjph.2022.06.017},
       adsurl = {https://ui.adsabs.harvard.edu/abs/2022ChJPh..78..141K}
}

@ARTICLE{2024EPJC...84..964J,
       author = {{Jumaniyozov}, Shokhzod and {Khan}, Saeed Ullah and {Rayimbaev}, Javlon and {Abdujabbarov}, Ahmadjon and {Urinbaev}, Sharofiddin and {Murodov}, Sardor},
        title = "{Circular motion and QPOs near black holes in Kalb{\textendash}Ramond gravity}",
      journal = {European Physical Journal C},
         year = 2024,
        month = sep,
       volume = {84},
       number = {9},
          eid = {964},
        pages = {964},
          doi = {10.1140/epjc/s10052-024-13351-y},
       adsurl = {https://ui.adsabs.harvard.edu/abs/2024EPJC...84..964J}
}

@ARTICLE{Will2014LRR....17....4W,
       author = {{Will}, Clifford M.},
        title = "{The Confrontation between General Relativity and Experiment}",
      journal = {Living Reviews in Relativity},
         year = 2014,
        month = dec,
       volume = {17},
       number = {1},
          eid = {4},
        pages = {4},
          doi = {10.12942/lrr-2014-4},
archivePrefix = {arXiv},
       eprint = {1403.7377},
 primaryClass = {gr-qc},
       adsurl = {https://ui.adsabs.harvard.edu/abs/2014LRR....17....4W}
}

@ARTICLE{Abbott2016PhRvL.116f1102A,
       author = {{Abbott}, B.~P. et al.},
        title = "{Observation of Gravitational Waves from a Binary Black Hole Merger}",
      journal = {\prl},
         year = 2016,
        month = feb,
       volume = {116},
       number = {6},
          eid = {061102},
        pages = {061102},
          doi = {10.1103/PhysRevLett.116.061102},
archivePrefix = {arXiv},
       eprint = {1602.03837},
 primaryClass = {gr-qc},
       adsurl = {https://ui.adsabs.harvard.edu/abs/2016PhRvL.116f1102A}
}

@ARTICLE{Bronnikov2022PhRvD.105d4039B,
       author = {{Bronnikov}, Kirill A. and {Walia}, Rahul Kumar},
        title = "{Field sources for Simpson-Visser spacetimes}",
      journal = {\prd},
         year = 2022,
        month = feb,
       volume = {105},
       number = {4},
          eid = {044039},
        pages = {044039},
          doi = {10.1103/PhysRevD.105.044039},
archivePrefix = {arXiv},
       eprint = {2112.13198},
 primaryClass = {gr-qc},
       adsurl = {https://ui.adsabs.harvard.edu/abs/2022PhRvD.105d4039B}
}

@ARTICLE{Remillard2002ApJ...580.1030R,
       author = {{Remillard}, Ronald A. and {Muno}, Michael P. and {McClintock}, Jeffrey E. and {Orosz}, Jerome A.},
        title = "{Evidence for Harmonic Relationships in the High-Frequency Quasi-periodic Oscillations of XTE J1550-564 and GRO J1655-40}",
      journal = {\apj},
         year = 2002,
        month = dec,
       volume = {580},
       number = {2},
        pages = {1030-1042},
          doi = {10.1086/343791},
archivePrefix = {arXiv},
       eprint = {astro-ph/0202305},
 primaryClass = {astro-ph},
       adsurl = {https://ui.adsabs.harvard.edu/abs/2002ApJ...580.1030R}
}

@ARTICLE{Remillard2006ARA&A..44...49R,
       author = {{Remillard}, Ronald A. and {McClintock}, Jeffrey E.},
        title = "{X-Ray Properties of Black-Hole Binaries}",
      journal = {\araa},
         year = 2006,
        month = sep,
       volume = {44},
       number = {1},
        pages = {49-92},
          doi = {10.1146/annurev.astro.44.051905.092532},
archivePrefix = {arXiv},
       eprint = {astro-ph/0606352},
 primaryClass = {astro-ph},
       adsurl = {https://ui.adsabs.harvard.edu/abs/2006ARA&A..44...49R}
}

@ARTICLE{Orosz2011ApJ...730...75O,
       author = {{Orosz}, Jerome A. and {Steiner}, James F. and {McClintock}, Jeffrey E. and {Torres}, Manuel A.~P. and {Remillard}, Ronald A. and {Bailyn}, Charles D. and {Miller}, Jon M.},
        title = "{An Improved Dynamical Model for the Microquasar XTE J1550-564}",
      journal = {\apj},
         year = 2011,
        month = apr,
       volume = {730},
       number = {2},
          eid = {75},
        pages = {75},
          doi = {10.1088/0004-637X/730/2/75},
archivePrefix = {arXiv},
       eprint = {1101.2499},
 primaryClass = {astro-ph.SR},
       adsurl = {https://ui.adsabs.harvard.edu/abs/2011ApJ...730...75O}
}

@ARTICLE{Motta2014MNRAS.437.2554M,
       author = {{Motta}, S.~E. and {Belloni}, T.~M. and {Stella}, L. and {Mu{\~n}oz-Darias}, T. and {Fender}, R.},
        title = "{Precise mass and spin measurements for a stellar-mass black hole through X-ray timing: the case of GRO J1655-40}",
      journal = {\mnras},
         year = 2014,
        month = jan,
       volume = {437},
       number = {3},
        pages = {2554-2565},
          doi = {10.1093/mnras/stt2068},
archivePrefix = {arXiv},
       eprint = {1309.3652},
 primaryClass = {astro-ph.HE},
       adsurl = {https://ui.adsabs.harvard.edu/abs/2014MNRAS.437.2554M}
}

@ARTICLE{Motta2014MNRAS.439L..65M,
       author = {{Motta}, S.~E. and {Munoz-Darias}, T. and {Sanna}, A. and {Fender}, R. and {Belloni}, T. and {Stella}, L.},
        title = "{Black hole spin measurements through the relativistic precession model: XTE J1550-564.}",
      journal = {\mnras},
         year = 2014,
        month = mar,
       volume = {439},
        pages = {L65-L69},
          doi = {10.1093/mnrasl/slt181},
archivePrefix = {arXiv},
       eprint = {1312.3114},
 primaryClass = {astro-ph.HE},
       adsurl = {https://ui.adsabs.harvard.edu/abs/2014MNRAS.439L..65M}
}

@ARTICLE{Ingram2019NewAR..8501524I,
       author = {{Ingram}, Adam R. and {Motta}, Sara E.},
        title = "{A review of quasi-periodic oscillations from black hole X-ray binaries: Observation and theory}",
      journal = {\nar},
         year = 2019,
        month = sep,
       volume = {85},
          eid = {101524},
        pages = {101524},
          doi = {10.1016/j.newar.2020.101524},
archivePrefix = {arXiv},
       eprint = {2001.08758},
 primaryClass = {astro-ph.HE},
       adsurl = {https://ui.adsabs.harvard.edu/abs/2019NewAR..8501524I}
}

@ARTICLE{Akaike1974ITAC...19..716A,
       author = {{Akaike}, H.},
        title = "{A New Look at the Statistical Model Identification}",
      journal = {IEEE Transactions on Automatic Control},
         year = 1974,
        month = jan,
       volume = {19},
        pages = {716-723},
          doi = {10.1109/TAC.1974.1100705},
       adsurl = {https://ui.adsabs.harvard.edu/abs/1974ITAC...19..716A}
}

@ARTICLE{2012JCAP...09..014B,
       author = {{Bambi}, Cosimo},
        title = "{Probing the space-time geometry around black hole candidates with the resonance models for high-frequency QPOs and comparison with the continuum-fitting method}",
      journal = {\jcap},
         year = 2012,
        month = sep,
       volume = {2012},
       number = {9},
          eid = {014},
        pages = {014},
          doi = {10.1088/1475-7516/2012/09/014},
archivePrefix = {arXiv},
       eprint = {1205.6348},
 primaryClass = {gr-qc},
       adsurl = {https://ui.adsabs.harvard.edu/abs/2012JCAP...09..014B}
}

@ARTICLE{2024CQGra..41a5032S,
       author = {{Saadati}, Reza and {Shojai}, Fatimah},
        title = "{Geodetic precession and shadow of quantum extended black holes}",
      journal = {Classical and Quantum Gravity},
         year = 2024,
        month = jan,
       volume = {41},
       number = {1},
          eid = {015032},
        pages = {015032},
          doi = {10.1088/1361-6382/ad13c3},
archivePrefix = {arXiv},
       eprint = {2312.12087},
 primaryClass = {gr-qc},
       adsurl = {https://ui.adsabs.harvard.edu/abs/2024CQGra..41a5032S}
}

@ARTICLE{2026JCAP...09..062R,
       author = {{Rodrigues}, Luiz F.~G. and {Lobo}, Francisco S.~N. and {Rodrigues}, Manuel E.},
        title = "{Lorentz-violating signatures in quasi-periodic oscillations from a magnetised Kalb-Ramond black hole}",
      journal = {\jcap},
         year = 2026,
        month = sep,
       volume = {2026},
       number = {9},
          eid = {062},
        pages = {062},
          doi = {10.1088/1475-7516/2026/09/062},
archivePrefix = {arXiv},
       eprint = {2606.03242},
 primaryClass = {gr-qc},
       adsurl = {https://ui.adsabs.harvard.edu/abs/2026JCAP...09..062R}
}

@ARTICLE{Pitkin2011LRR.14..5P,
       author = {{Pitkin}, Matthew and {Reid}, Stuart and {Rowan}, Sheila and {Hough}, James},
        title = "{Gravitational Wave Detection by Interferometry (Ground and Space)}",
      journal = {Living Reviews in Relativity},
         year = 2011,
        month = jul,
       volume = {14},
       number = {1},
          eid = {5},
        pages = {5},
          doi = {10.12942/lrr-2011-5},
archivePrefix = {arXiv},
       eprint = {1102.3355},
 primaryClass = {astro-ph.IM},
       adsurl = {https://ui.adsabs.harvard.edu/abs/2011LRR....14....5P}
}

@ARTICLE{Liu2024PhRvD.109h4074L,
       author = {{Liu}, Tonghua and {Biesiada}, Marek and {Tian}, Shuxun and {Liao}, Kai},
        title = "{Robust test of general relativity at the galactic scales by combining strong lensing systems and gravitational wave standard sirens}",
      journal = {\prd},
         year = 2024,
        month = apr,
       volume = {109},
       number = {8},
          eid = {084074},
        pages = {084074},
          doi = {10.1103/PhysRevD.109.084074},
archivePrefix = {arXiv},
       eprint = {2404.05907},
 primaryClass = {gr-qc},
       adsurl = {https://ui.adsabs.harvard.edu/abs/2024PhRvD.109h4074L}
}

@ARTICLE{Li2026arXiv260214458L,
       author = {{Li}, Ziyang and {Liu}, Shou-Qi and {Huang}, Jia-Hui},
        title = "{Circular orbits and observational features of the rotating Simpson-Visser black hole surrounded by a thin accretion disk}",
      journal = {arXiv e-prints},
         year = 2026,
        month = feb,
          eid = {arXiv:2602.14458},
        pages = {arXiv:2602.14458},
          doi = {10.48550/arXiv.2602.14458},
archivePrefix = {arXiv},
       eprint = {2602.14458},
 primaryClass = {astro-ph.HE},
       adsurl = {https://ui.adsabs.harvard.edu/abs/2026arXiv260214458L}
}

@ARTICLE{Strohmayer_2001ApJ...552L..49S,
       author = {{Strohmayer}, Tod E.},
        title = "{Discovery of a 450 HZ Quasi-periodic Oscillation from the Microquasar GRO J1655-40 with the Rossi X-Ray Timing Explorer}",
      journal = {\apjl},
         year = 2001,
        month = may,
       volume = {552},
       number = {1},
        pages = {L49-L53},
          doi = {10.1086/320258},
       adsurl = {https://ui.adsabs.harvard.edu/abs/2001ApJ...552L..49S}
}

@ARTICLE{2015EPJC...75..162B,
       author = {{Bambi}, Cosimo},
        title = "{Testing the nature of the black hole candidate in GRO J1655-40 with the relativistic precession model}",
      journal = {European Physical Journal C},
         year = 2015,
        month = apr,
       volume = {75},
          eid = {162},
        pages = {162},
          doi = {10.1140/epjc/s10052-015-3396-7},
archivePrefix = {arXiv},
       eprint = {1312.2228},
 primaryClass = {gr-qc},
       adsurl = {https://ui.adsabs.harvard.edu/abs/2015EPJC...75..162B}
}

@ARTICLE{Beer2002MNRAS.331..351B,
       author = {{Beer}, Martin E. and {Podsiadlowski}, Philipp},
        title = "{The quiescent light curve and the evolutionary state of GRO J1655-40}",
      journal = {\mnras},
         year = 2002,
        month = mar,
       volume = {331},
       number = {2},
        pages = {351-360},
          doi = {10.1046/j.1365-8711.2002.05189.x},
archivePrefix = {arXiv},
       eprint = {astro-ph/0109136},
 primaryClass = {astro-ph},
       adsurl = {https://ui.adsabs.harvard.edu/abs/2002MNRAS.331..351B}
}

@ARTICLE{2023MNRAS.523..375S,
       author = {{Shaikh}, Rajibul},
        title = "{Testing black hole mimickers with the Event Horizon Telescope image of Sagittarius A*}",
      journal = {\mnras},
         year = 2023,
        month = jul,
       volume = {523},
       number = {1},
        pages = {375-384},
          doi = {10.1093/mnras/stad1383},
archivePrefix = {arXiv},
       eprint = {2208.01995},
 primaryClass = {gr-qc},
       adsurl = {https://ui.adsabs.harvard.edu/abs/2023MNRAS.523..375S}
}

@ARTICLE{Rezzolla2014PhRvD,
       author = {{Rezzolla}, Luciano and {Zhidenko}, Alexander},
        title = "{New parametrization for spherically symmetric black holes in metric theories of gravity}",
      journal = {\prd},
         year = 2014,
        month = oct,
       volume = {90},
       number = {8},
          eid = {084009},
        pages = {084009},
          doi = {10.1103/PhysRevD.90.084009},
archivePrefix = {arXiv},
       eprint = {1407.3086},
 primaryClass = {gr-qc},
       adsurl = {https://ui.adsabs.harvard.edu/abs/2014PhRvD..90h4009R}
}

@ARTICLE{Konoplya2020PhRvD.101l4004K,
       author = {{Konoplya}, R.~A. and {Zhidenko}, A.},
        title = "{General parametrization of black holes: The only parameters that matter}",
      journal = {\prd},
         year = 2020,
        month = jun,
       volume = {101},
       number = {12},
          eid = {124004},
        pages = {124004},
          doi = {10.1103/PhysRevD.101.124004},
archivePrefix = {arXiv},
       eprint = {2001.06100},
 primaryClass = {gr-qc},
       adsurl = {https://ui.adsabs.harvard.edu/abs/2020PhRvD.101l4004K}
}

@ARTICLE{Jumaniyozov2024EPJC964y,
       author = {{Jumaniyozov}, Shokhzod and {Khan}, Saeed Ullah and {Rayimbaev}, Javlon and {Abdujabbarov}, Ahmadjon and  {Urinbaev}, Sharofiddin and {Murodov}, Sardor}, 
        title = "{Circular motion and QPOs near black holes in Kalb–Ramond gravity}",
      journal = {European Physical Journal C},
         year = 2024,
        month = mar,
       volume = {84},
       number = {964},
        doi = {10.1140/epjc/s10052-024-13351-y}
}

@ARTICLE{2026EPJC...86..510S,
       author = {{Sui}, Tao-Tao and {Long}, Chen and {Zhang}, Ye},
        title = "{The properties and predictions of quasi-periodic oscillations around a black hole in nonlocal gravity}",
      journal = {European Physical Journal C},
         year = 2026,
        month = may,
       volume = {86},
       number = {5},
          eid = {510},
        pages = {510},
          doi = {10.1140/epjc/s10052-026-15770-5},
       adsurl = {https://ui.adsabs.harvard.edu/abs/2026EPJC...86..510S}
}

@ARTICLE{2012MNRAS.426.1701B,
       author = {{Belloni}, T.~M. and {Sanna}, A. and {M{\'e}ndez}, M.},
        title = "{High-frequency quasi-periodic oscillations in black hole binaries}",
      journal = {\mnras},
         year = 2012,
        month = nov,
       volume = {426},
       number = {3},
        pages = {1701-1709},
          doi = {10.1111/j.1365-2966.2012.21634.x},
archivePrefix = {arXiv},
       eprint = {1207.2311},
 primaryClass = {astro-ph.HE},
       adsurl = {https://ui.adsabs.harvard.edu/abs/2012MNRAS.426.1701B}
}

@ARTICLE{2001A&A...374L..19A,
       author = {{Abramowicz}, M.~A. and {Klu{\'z}niak}, W.},
        title = "{A precise determination of black hole spin in GRO J1655-40}",
      journal = {\aap},
         year = 2001,
        month = aug,
       volume = {374},
        pages = {L19-L20},
          doi = {10.1051/0004-6361:20010791},
archivePrefix = {arXiv},
       eprint = {astro-ph/0105077},
 primaryClass = {astro-ph},
       adsurl = {https://ui.adsabs.harvard.edu/abs/2001A&A...374L..19A}
}

@ARTICLE{1998ApJ...499L..37M,
       author = {{Miller}, M. Coleman and {Lamb}, Frederick K.},
        title = "{Bounds on the Compactness of Neutron Stars from Brightness Oscillations during X-Ray Bursts}",
      journal = {\apjl},
         year = 1998,
        month = may,
       volume = {499},
       number = {1},
        pages = {L37-L40},
          doi = {10.1086/311335},
archivePrefix = {arXiv},
       eprint = {astro-ph/9711325},
 primaryClass = {astro-ph},
       adsurl = {https://ui.adsabs.harvard.edu/abs/1998ApJ...499L..37M}
}

@ARTICLE{1998ApJ...509L..37K,
       author = {{Klu{\'z}niak}, W.},
        title = "{General Relativistic Constraints on the Equation of State of Dense Matter Implied by Kilohertz Quasi-periodic Oscillations in Neutron-Star X-Ray Binaries}",
      journal = {\apjl},
         year = 1998,
        month = dec,
       volume = {509},
       number = {1},
        pages = {L37-L40},
          doi = {10.1086/311748},
archivePrefix = {arXiv},
       eprint = {astro-ph/9712243},
 primaryClass = {astro-ph},
       adsurl = {https://ui.adsabs.harvard.edu/abs/1998ApJ...509L..37K}
}

@ARTICLE{2022MPLA...3750064K,
       author = {{Khan}, Saeed Ullah and {Ren}, Jingli and {Rayimbaev}, Javlon},
        title = "{Circular motion around a regular rotating Hayward black hole}",
      journal = {Modern Physics Letters A},
         year = 2022,
        month = apr,
       volume = {37},
       number = {11},
          eid = {2250064},
        pages = {2250064},
          doi = {10.1142/S021773232250064X},
       adsurl = {https://ui.adsabs.harvard.edu/abs/2022MPLA...3750064K}
}

@ARTICLE{1993ARA&A..31...93V,
       author = {{Verbunt}, Frank},
        title = "{Origin and evolution of X-ray binaries and binary radio pulsars.}",
      journal = {\araa},
         year = 1993,
        month = jan,
       volume = {31},
        pages = {93-127},
          doi = {10.1146/annurev.aa.31.090193.000521},
       adsurl = {https://ui.adsabs.harvard.edu/abs/1993ARA&A..31...93V}
}

@ARTICLE{2022CQGra..39g5021R,
       author = {{Rayimbaev}, Javlon and {Bokhari}, Ashfaque Hussain and {Ahmedov}, Bobomurat},
        title = "{Quasiperiodic oscillations from noncommutative inspired black holes}",
      journal = {Classical and Quantum Gravity},
         year = 2022,
        month = apr,
       volume = {39},
       number = {7},
          eid = {075021},
        pages = {075021},
          doi = {10.1088/1361-6382/ac556a},
       adsurl = {https://ui.adsabs.harvard.edu/abs/2022CQGra..39g5021R}
}

@ARTICLE{Stuchlik2021JCAP...11..059S,
       author = {{Stuchl{\'\i}k}, Z. and {Vrba}, J.},
        title = "{Supermassive black holes surrounded by dark matter modeled as anisotropic fluid: epicyclic oscillations and their fitting to observed QPOs}",
      journal = {\jcap},
         year = 2021,
        month = nov,
       volume = {2021},
       number = {11},
          eid = {059},
        pages = {059},
          doi = {10.1088/1475-7516/2021/11/059},
archivePrefix = {arXiv},
       eprint = {2110.07411},
 primaryClass = {gr-qc},
       adsurl = {https://ui.adsabs.harvard.edu/abs/2021JCAP...11..059S}
}

@ARTICLE{2021JCAP...07..036F,
       author = {{Franzin}, Edgardo and {Liberati}, Stefano and {Mazza}, Jacopo and {Simpson}, Alex and {Visser}, Matt},
        title = "{Charged black-bounce spacetimes}",
      journal = {\jcap},
         year = 2021,
        month = jul,
       volume = {2021},
       number = {7},
          eid = {036},
        pages = {036},
          doi = {10.1088/1475-7516/2021/07/036},
archivePrefix = {arXiv},
       eprint = {2104.11376},
 primaryClass = {gr-qc},
       adsurl = {https://ui.adsabs.harvard.edu/abs/2021JCAP...07..036F}
}

\end{document}